# Solvation and thermalization of electrons generated by above the gap (12.4 eV) two-photon ionization of liquid $H_2O$ and $D_2O$. [1]

Rui Lian, Robert A. Crowell, [*] and Ilya A. Shkrob.

[a] *Chemistry Division, Argonne National Laboratory, Argonne, IL 60439*

**Abstract**

Temporal evolution of transient absorption (TA) spectra of electrons generated by above-the-gap (12.4 eV total energy) two-photon ionization of liquid $H_2O$ and $D_2O$ has been studied on femto- and pico- second time scales. The spectra were obtained at intervals of 50 nm between 0.5 and 1.7 µm. Two distinct regimes of the spectral evolution were observed: $t < 1$ ps and $t > 1$ ps. In both of these regimes, the spectral profile changes considerably with the delay time of the probe pulse. The "continuous blue shift" and the "temperature jump" models, in which the spectral profile does not change as it progressively shifts, as a whole, to the blue, are not supported by our data. Furthermore, no *p*-state electron, postulated by several authors to be a short-lived intermediate of the photoionization process, was observed by the end of the 300 fs, 200 nm pump pulse.

For $t < 1$ ps, two new TA features (the 1.15 µm band and 1.4 µm shoulder) were observed for the electron in the spectral region where O-H overtones appear in the spectra of light water. These two features were not observed for the electron in $D_2O$. Vibronic coupling to the modes of water molecules lining the solvation cavity is a possible origin of these features. On the sub-picosecond time scale, the absorption band of solvated electron progressively shifts to the blue. At later delay times ($t > 1$ps), the position of the band maximum is "locked", but the spectral profile continues to change by narrowing on the red side and broadening on the blue side; the oscillator strength is constant within 10%. The time constant of this narrowing is ca. 0.56 ps for $H_2O$ and 0.64 ps for $D_2O$. Vibrational relaxation and time-dependent decrease in the size and sphericity of the solvation cavity are suggested as possible causes for the observed spectral transformations in both of these regimes.

______________________________________________________

[*]  To whom correspondence should be addressed: *Tel* 630-252-8089, *FAX* 630-2524993, *e-mail:* rob_crowell@anl.gov.

1. Introduction.

Although ultrafast dynamics of the excess electron in liquid water have been extensively studied over the last two decades, many questions concerning the mechanism of localization, solvation and thermalization of this electron still remain. Using femtosecond pump-probe transient absorbance (TA) laser spectroscopy, short-lived precursors (which in the following will be collectively called "pre-solvated" electrons) of the hydrated, thermalized electron have been observed on sub-picosecond to picosecond time scales. [1-17] Common means of generating these pre-solvated electrons include multiphoton ionization of neat water, [e.g., 1-6,13-16] photoionization of molecular solutes and electron photodetachment from aqueous anions, [e.g., 2,16,17] and photoexcitation of hydrated electrons. [e.g., 7-12] The initial studies [1-5] were limited to the observation of the TA signal for a few strategically chosen probe wavelengths in the visible and near-infrared (> 1 µm). Recently, the observation range has been expanded into the ultraviolet (out to 200 nm) [14] and mid-infrared (out to 5 µm). [16] The latter studies seem to suggest the existence of two infrared (IR) absorbing precursors of the hydrated electron, $e_{aq}^-$. One of these species has a broad absorption band centered at 1.6 µm (suggesting a localized state in a shallow trap), [16] while another species has a featureless spectrum that resembles that of a free (extended state) conduction band electron in semiconductors or a long-lived, IR-absorbing electron in low-temperature ice. [19-21] While the existence of the short-lived IR-absorbing species has been hinted at by the pioneering studies of Migus et al. [1] and spectral decomposition analyses of Gauduel and co-workers, [3-5] subsequent work on electron solvation in photoionization of water favored the so-called "continuous shift" model [e.g., 6,13,17,22-25] in which the existence of these IR-absorbing species (such as shallow-trap "wet" electrons) [e.g., 6,11,16] is not strictly required. In the latter model, the absorption profile $S(E,t)$ of the pre-solvated electron as a function of photon energy $E$ is postulated to shift, as a whole, to the blue during the solvation, without changing its shape.

In a more recent variant of this model suggested by Madsen et al. [15] (which is referred to in the following as the "temperature jump" model), the thermalization is regarded as "cooling" of water around the solvation cavity in which the absorption spectrum of the electron at any time during the thermalization process is identical to the spectrum of hydrated electron in the state of equilibrium with the solvent at some higher temperature. In this model, the electron thermalization is viewed as a succession of quasi-equilibrium states that are fully characterized by the time evolution of the local temperature. This model, like the original "continuous shift" model is purely phenomenological: no explanation is given on theoretical grounds, as to why such a picture of the electron solvation might be correct. Since the width of the TA spectrum for $e_{aq}^-$ depends weakly on the water temperature, whereas the position of the absorption maximum is strongly temperature dependent, [26] from the standpoint of data analysis, the "temperature jump" model of Madsen et al. [15] is nearly identical to the "continuous shift" model. Despite their appealing simplicity, both of these models are inadequately supported.

First, in many ultrafast studies, the TA spectra were too sparsely sampled: the TA kinetics were obtained for just 4-to-10 wavelengths of the probe light. [e.g., 3-5,13,17] This sparse sampling makes it impossible to verify the central tenet of the "continuous shift" model, *viz*. the constancy of the spectral profile during the thermalization process. Typically, this constancy was *postulated* rather than observed: For a given delay time *t* of the probe pulse, the (sparse) TA spectrum was fit by a constant-profile template using least squares optimization. As a result, the experimental spectrum is characterized by a single parameter, $E_{max}(t)$, the photon energy corresponding to the absorption maximum. Naturally, this procedure tells little about the robustness of the template used to fit the experimental spectrum since any deviation from the prescribed shape is regarded as statistical deviation. Worse, the exact location of the maximum strongly depends on the template used to fit the data, inasmuch as the electron spectra are flat at the top, broad, and featureless, so that $E_{max}(t)$ is determined, implicitly or explicitly, by the extrapolation of the data in the spectral wings towards the center, where the TA signal is maximum. If the profile of the sparsely-sampled TA spectra were time-dependent (as suggested by our results), this approach would yield grossly incorrect results.

Second, the TA studies in which the spectral sampling was sufficiently dense (at least 10-15 wavelengths of the probe light across the spectrum) *did not* fully support the "continuous shift" model. E.g., the benchmark studies by Gauduel and co-workers [3-5] and

4.

Pepin et al. [6] on electron dynamics in $H_2O$ and $D_2O$, respectively, suggested that the "continuous shift" alone cannot account for the observed spectral evolution. In the latter study, an additional long-lived (> 2 ps) species with a spectral band centered at 1.1-1.3 µm was postulated to explain the evolution of the TA spectrum for $\lambda$ >0.9 µm. Subsequent pump-probe studies failed to verify the existence of such a long-lived species. Part of the problem was that the temporal evolution of spectrum in the near-infrared was obtained using TA kinetics that exhibited poor signal-to-noise (S/N) ratio (3:1 to 10:1). Thus, Pepin et al. [6] conceded that more TA studies with dense sampling of the TA kinetics across a wider spectral region and better-quality kinetic data would be needed to characterize the spectral evolution and establish the validity of the "continuous shift" model. The present study implements such a program.

Third, it is now understood that the electrons generated in different photoprocesses have somewhat different thermalization [e.g., 9] and geminate recombination [9,27-29] dynamics. Specifically, in photoionization of neat water, different excitation mechanisms operate for photons of different energy. [27-29] Two- or three-photon excitation at the lower photon energy (so that the total excitation energy < 9 eV) produces electrons that are, on average, just 0.9-1 nm away from their parent hole, whereas at higher total photon energy (> 11 eV), the photoelectrons are at least 2-3 nm away. [27,28] It is believed that for high-energy photoexcitation, the electrons are ejected directly into the conduction band of the solvent; at lower energy, the photoionization involves concerted proton and electron transfers, perhaps to pre-existing traps. [30-32] In the intermediate regime, autoionization of water is thought to compete with these two photoprocesses. [28] The studies with the densest spectral sampling [e.g., 5,6] were carried out in these low-energy and intermediate regimes. Due to the short separation distances between the geminate partners, some electrons decayed during the fist few picoseconds when their solvation and thermalization had occurred. [6] This decay had to be taken into account, by extrapolation of the picosecond dynamics to short delay times. Such a procedure implies that the pre-solvated electron follows the same recombination dynamics as fully hydrated and thermalized electron, which may not be correct. E.g., recent simulations of spectral evolution for electrons generated by photodetachment from aqueous iodide (carried out using the "continuous shift" model) by Vilchiz et al. [17] suggested significant recombination of the short-lived pre-solvated species.

In the present study, femto- and pico- second TA kinetics were obtained for the electron generated by a high-energy ionization photoprocess (12.4 eV total energy); in this regime, geminate recombination on the picosecond time scale is unimportant. [29] Short (300 fs fwhm) 200 nm laser pulses were used to photoionize room temperature $H_2O$ and $D_2O$ via biphotonic excitation. A set of TA kinetics obtained was sampled out to 5-30 ps in steps of 50 nm from 0.5 to 1.4 µm (to 1.7 µm for $D_2O$). Throughout this entire spectral range, the S/N ratio for the individual kinetics was better than 50:1.

Our results suggest that the "continuous shift" model does not fully account for the spectral evolution occurring during electron solvation/thermalization in light and heavy water. The shape of the TA spectra is not constant, and the thermalization dynamics cannot be reduced to the shift of the band maximum alone. For $t > 1$ ps, the electron spectrum exhibits progressive line narrowing in the red (concomitant with the line broadening in the blue). This change continues well after the spectral shift is completed; it probably results from the vibrational relaxation of water molecules around the "hot" ground-state electron. Furthermore, for the electron in $H_2O$, new features were observed in the near-infrared on the sub-picosecond time scale. One of these features might have been observed previously by Laenen et al. [16] in $H_2O$ and $D_2O$ (a band centered at 1.6 µm). Another, with a band centered at 1.15 µm, might be the tentative IR-absorbing species postulated by Jay-Gerin and coworkers. [6,23,24] In the present study, no evidence of these two features has been found in $D_2O$, in a striking contradiction to the results of Laenen et al. [16] It appears that these absorption bands do not originate from a purely electronic transition. Rather, these bands seem to involve *vibrations* of the water molecules.

We forewarn the reader that no specific model that accounts for these observations is given in this paper. Many models of electron solvation and relaxation, at all levels of theory, have been suggested. Given the limited success that these theories met in explanation of experimental observations, offering yet another such model seems injudicious. Rather, we focus on the spectral features themselves and what can be glimpsed from these features without the benefit of theoretical insight. Such an insight is certainly needed, but it is beyond the scope of this study.

Hereafter, the index "S" (e.g., Fig. 1S) indicates that the material is placed in the Supporting Information available electronically from the journal. Appendices A and B are also placed therein.

**2. Experimental.**

*Materials.* Deionized water with conductivity < 2 nS/cm was used in all experiments with $H_2O$. An $N_2$-saturated 1 L sample was circulated using a gear pump through a jet nozzle. A 500 mL sample of heavy water (99 atom %, Aldrich) was used in all experiments with $D_2O$. No change in the kinetics was observed after extensive photolysis of this sample. The details of the flow system are given elsewhere. [33]

*Ultrafast laser spectroscopy.* The pico- and femto- second TA measurements were carried out using a 1 kHz Ti:sapphire setup, details of which are given in refs. 29 and 33. This setup provided 60 fs FWHM, 3 mJ light pulses centered at 800 nm. One part of the beam was used to generate probe pulses while the other part was used to generate the 200 nm (fourth harmonic) pump pulses. Up to 20 μJ of the 200 nm light was produced this way (300-350 fs FWHM pulse). The pump and probe beams were perpendicularly polarized and overlapped at the surface of a 90 μm thick high-speed water jet at $5^o$. No change of the TA kinetics with the polarization of the probe light was observed.

A white light supercontinuum was generated by focussing the 800 nm fundamental on a 1 mm thick sapphire disk; the probe light was selected using a set of 10 nm FWHM interference filters (Corion). For two wavelengths, 1.6 and 1.7 μm (used in the studies of electron in $D_2O$), the intensity of the probe light was weak and 30 nm FWHM bandpass filters were used instead (which resulted in a shorter probe pulse). Appropriate glass cutoff filters were used to block stray 200, 400, and 800 nm light. For measurements in the near-infrared, $0^o$ dielectric mirrors for 800 nm light were inserted in the path of the probe beam to reduce the leakage of the fundamental. Fast Si photodiodes (FND100Q from EG&G) biased at -90 V were used for detection of the $\lambda<1.3$ μm light, fast Ge photodiodes (GMP566 from GPD Optoelectronics Corp.) biased at -9 V were used for detection of $\lambda > 1.2$ µm. Due to the considerable dark current for the Ge photodiodes, the load resistance was relatively low, ca. 0.2 MΩ vs. 6 MΩ for Si photodiode, which led to lower sensitivity. The vertical bars in the figures represent 95% confidence limits for each data point.

To study the thermalization dynamics of the electron, the delay time was increased in steps of 50 fs to 5-7 ps and steps of 300 fs to 25 ps. The kinetic origin (i.e., zero delay time $t$) and the $1/e$ width $\tau_p$ of the 200 nm excitation pulse convoluted with the probe pulse were determined by following the TA signal from a 1.4 µm thick amorphous Si:H alloy (8 atom % H) film on a suprasil substrate.[34] The instantaneous increase in the TA signal near the kinetic origin is from the generation of free electron carriers in the Si:H sample,[35] the slower decay kinetics are due to the thermalization, trapping, and recombination of these photocarriers.[34,35] Deconvolution of these TA signals (represented as a Gaussian pulse convoluted with biexponential decay to a plateau) gives the "pulse width" $\tau_p$ (characteristic of the response function of the setup). Due to the considerable chirp in the 200 nm pump pulse (introduced by harmonic-generating BBO crystals and focussing optics), $\tau_p$ was 190-290 fs, depending on the optimization of the compressor and the probe wavelength. The latter dependence is due to the wavelength-dependent chirp in the white light continuum and group velocity mismatch (GVM) in the water sample (traceable to the difference in the speed of light for the pump and probe pulses). The variation in the pulse width and the resulting uncertainty in the time origin made it difficult to obtain good-quality electron spectra for $t < 500$ fs. For $\lambda < 700$ nm, the TA traces exhibit a sharp "spike" at the time origin (see Fig. 1(a) in section 3) whose time profile follows the photoresponse function of the detection system. This TA signal in the visible has been observed by other workers [e.g., 6,13,17] and originates from nonlinear absorbance of the excitation and probe light in the sample. It can be used to juxtapose the corresponding kinetics in time without a reference to the Si:H sample. Since the photoexcitation wavelength of 200 nm is right at the onset of one-photon water ionization (the quantum yield of this ionization is ca. $1.6 \times 10^{-2}$ at 193 nm),[36] this artifact may be present even when red and near-infrared light is used to probe the electron dynamics (section 3).

The kinetic traces given below were obtained using a 1-5 µJ pump pulse focussed, using a thin $MgF_2$ lens, to a round spot of 300 µm FWHM; the probe beam was typically 50-60 µm FWHM. The typical TA signal ($\Delta OD_\lambda$, where $\lambda$ is the wavelength of the probe light in nanometers) at the maximum was 10-to-50 mOD. The TA signal $\Delta OD_{800}(t)$ from hydrated electron obtained at the delay time $t$=10 ps plotted as a function of 200 nm power scaled as the square of the pulse energy, indicating biphotonic ionization.[29]

*Reconstruction of the electron spectrum.* Time-dependent TA spectra $S(\lambda,t)$ given below were obtained from the TA kinetics using the following approach:[13,17] The kinetics (obtained independently for each probe wavelength $\lambda$) were normalized by the average of the TA signal at $t$=5-8 ps. At this delay time, the $\Delta OD_\lambda(t)$ kinetics reach a plateau: the thermalization phase is complete (see section 3) yet the electron decay due to geminate recombination is still negligible.[29] These normalized kinetics were then weighted by the spectrum $S(\lambda)$ of the thermalized (hydrated electron) to obtain time-dependent spectrum $S(\lambda,t)$ of the pre-solvated electron:

$$S(\lambda,t) = S(\lambda) \; \Delta OD_\lambda(t)/\Delta OD_\lambda(t = 5-8 \; ps) \qquad (1)$$

For $\lambda > 1$ µm, where $e_{aq}^-$ absorbs poorly, the use of this procedure requires more comment (see Appendix A).

First, since the TA signal from the hydrated electron at $t$>5 ps becomes very small relative to the "spike" from the pre-solvated electron (Figs. 1, 2, and 3), even a small error of a few per cent in the scaling factor $S(\lambda)$ in eq. (1) has a large effect on the extrapolated $S(\lambda,t)$ signal at short delay time. Thus, a high-quality near-infrared spectrum of $e_{aq}^-$ in light and heavy water was needed. To this end, a separate flash photolysis study was carried out to obtain a spectrum of $e_{aq}^-$ in $H_2O$ ($\lambda$=0.4-1.4 µm) and $D_2O$ ($\lambda$=0.4-1.7 µm) (see Appendix A.1 and A.2). Second, one should be aware that the weak, long-lived TA signal observed at $t > 5$ ps might not be related to $e_{aq}^-$ since the solvent itself weakly absorbs in the near-infrared (where overtones of the O-H stretch appear). Specifically, a long-lived thermal signal from laser-heated water could change the transmittance of the probe light. To address such a possibility, 1 M HClO$_4$ was added to H$_2$O (hydronium ion scavenges $e_{aq}^-$ with rate constant of 2.3x10$^{10}$ M$^{-1}$ s$^{-1}$).[27] This scavenging removed most of the TA signal in the first 300 ps. By comparing the long-term kinetic traces obtained at different wavelengths, we were able to demonstrate that at least 90-95% of the TA signal at 1.1-1.3 µm was from $e_{aq}^-$ (Fig. 4S in the Supplement). However, for $\lambda > 1.3$ µm, the long-lived TA signal was too weak to obtain good-quality decay kinetics on the sub-nanosecond time scale. Thus, we cannot presently confirm that *all* of the long-lived TA signal in the near-infrared is due to the absorbance from $e_{aq}^-$. This uncertainty introduces ambiguity in the reconstruction of the near-infrared spectra using eq. (1). We will return to this point in section 3.

Third, it is nearly impossible to retain the same characteristics of the probe light (such as its pulse width, chirp, and GVM) across the entire spectral region of interest. For this reason, obtaining reliable TA spectra $S(\lambda,t)$ using eq. (1) for delay times $t < 500$ fs is impossible without (i) making specific assumptions about these pulse characteristics and (ii) using *ad hoc* kinetic schemes for spectrum modeling (as done, for example, in refs. 5, 16, 17, 22, and 23). Since these kinetic schemes are not unique, the robustness of such analyses is impossible to estimate. Our approach is to analyze the data with as few *ad hoc* assumptions as possible, and for this reason we focus on the evolution of the TA spectra that occurs between 0.5 and 3 ps after the 12.4 eV photoionization.

## 3. Results.

Typical TA kinetics observed for pre-solvated electron following two 200 nm photon ionization of light water are shown in Figs. 1, 2, and 3. The complete set of digitized kinetic traces is given in the supplement (see Appendix B for instructions). Most of the trends observed in these kinetic traces have already been seen by others. For $\lambda > 0.8$ µm, the TA signal rapidly increases within the duration of the 200 nm pulse and then slowly decreases to a plateau (Figs. 1(b), 2(b), and 3(a)). The ratio of the maximum TA signal to the signal attained at $t=5$ ps increases with increasing probe wavelength $\lambda$ (e.g., Fig. 3). For D$_2$O, this ratio increases from ca. 5:1 at 1.1 µm to ca. 10:1 at 1.25 µm to ca. 30:1 at 1.4 µm to ca. 50:1 at 1.6 µm. As the plateau absorbance becomes smaller (with respect to the maximum TA signal) with the increasing $\lambda$, the decay kinetics also become progressively faster (e.g., compare the 0.85 and 1.5 µm traces in Figs. 2(a) and 3(a), respectively). In the visible ($\lambda=0.5$-0.75 µm), the increase in the TA signal is relatively slow (which occurs over the first 2 ps after the ionization, Figs. 1(a) and 2(a)) and the changes in the kinetic profiles with the wavelength are less pronounced than for $\lambda > 0.8$ µm.

Qualitatively, these trends can be accounted for by a progressive blue shift of the absorption band of the ground state electron in the course of its solvation and thermalization. [6,13,17,22-24] Initially, the electron localizes in a large cavity and interacts with the H-O dipoles of water molecules weakly, absorbing in the near-infrared. At later delay times, the cavity contracts, the interaction becomes stronger, and the absorption band shifts to the visible so that the TA signal in the near-infrared rapidly decays (since $e_{aq}^-$ is a weak absorber in this region) and the TA signal in the visible increases (since $e_{aq}^-$ absorbs strongly in the red). The ratio of the maximum TA signal to the plateau value increases with the wavelength $\lambda$ because the absorptivity of $e_{aq}^-$ rapidly decreases in the same direction. The change in the TA signal is faster where a small change in the wavelength results in a large change in the TA signal, as occurs in the near-infrared. In this general outline of the spectral evolution, our study fully concurs with the previous results. However, these features *per se* do not imply that the profile of the TA spectrum remains constant during the spectral shift.

To characterize the spectral evolution of photoelectron during its solvation, the method outlined in section 2 and Appendix A has been used. For reasons explained therein, the use of this method is justified only for relatively long delay times, $t > 0.5$ ps. The spectra $S(\lambda,t)$ obtained using eq. (1) were linearly interpolated in the time domain and then integrated between the delay times $t_1$ and $t_2 = t_1 + \Delta t$, to obtain the average

11.

spectrum for the corresponding $(t_1,t_2)$ window. Figs. 4(a) and 5(a) show several such spectra for electrons in light and heavy water, respectively, for delay times $t_1$ between 0.5 and 1.2 ps ($\Delta t$=100 fs). Figs. 4(b) and 5(b) show the spectral evolution at later delay times, from 1.3 to 2.7 ps ($\Delta t$=200 fs). For comparison, the spectrum of a fully thermalized, solvated electron in $H_2O$ and $D_2O$ is indicated by a bold line. For $t >$ 1ps, the spectral evolution of the electron in light water (Fig. 4(b)) is very similar to that in heavy water (Fig. 5(b)). By contrast, on the sub-picosecond time scale, near-infrared spectra for electron in $H_2O$ (Fig. 4(a)) look quite different from those in $D_2O$ (Fig. 5(a)). See also Fig. 5S where the data of Figs. 4(a) and 5(a) are plotted vs. the photon energy, which emphasizes the systematic change in the shape of the energy profile with delay time.

On the short time scale (0.5-1.2 ps), the absorption band of the excess electron rapidly shifts to the blue. As the band shifts, the spectrum narrows on both sides of the maximum (Fig. 5S). At $t$=1-1.2 ps, the position $E_{max}$ of the absorption maximum almost reaches its equilibrium value. For $t >$ 1.2 ps, the band maximum is "locked", however, the spectral evolution continues: on the red side, the absorption line becomes narrower and on the blue side, it becomes broader (Figs. 4(b) and 5(b)). Thus, there are two distinct regimes in the spectral evolution of the electron: (i) the "fast (sub-picosecond) regime", in which the spectral features are dominated by the band shift to the blue and overall spectral narrowing and (ii) the "slow (> 1 ps) regime", in which the spectral evolution is dominated by small-scale spectral transformations occurring after $E_{max}$ reaches its equilibrium value.

To better characterize the latter regime, the experimental TA spectra were fit to the same Lorentzian-Gaussian curves used to simulate the spectrum of hydrated electron: 26,37,38

$$S(E) = \left(1 + \left[(E - E_{max})/W_L\right]^\nu\right)^{-1} \quad for \ \ E > E_{max}, \tag{2}$$

$$S(E) = \exp\left(-\left[(E - E_{max})/W_G\right]^2\right) \quad for \ \ E < E_{max} \tag{3}$$

where $E$ is the photon energy (the spectrum $S(E)$ is normalized at the band maximum at $E = E_{max}$), $v=2$ is the exponent, and $W_L$ and $W_G$ are the Lorentzian and Gaussian widths, respectively. Each spectrum $S(E,t)$ averaged over the $(t_1, t_1 + \Delta t)$ window has been fit to eqs. (2) and (3) using least squares optimization (with a floating scaling factor), and the optimum parameters $E_{max}$, $W_L$, and $W_G$ are plotted in Fig. 6 vs. the delay time $t_1$ ($\Delta t=30$ fs slices).

It is obvious from these plots that the increase in the band energy $E_{max}$ is much faster than the decrease in the Gaussian width $W_G$ (compare Figs. 6(a) and 6(b)). For $t > 0.5$ ps, both of these parameters exponentially approach their respective equilibrium values $E_{max}^\infty$ and $W_G^\infty$. For $E_{max}$, the corresponding first order rate constant is ca. 3.1 ps$^{-1}$ (which compares well with other estimates; see, e.g., refs. 3-9, 16, and 17). For the Gaussian width $W_G$, the rate constants are 1.62±0.4 ps$^{-1}$ for H$_2$O and 1.45±0.2 ps$^{-1}$ for D$_2$O. It is this large difference in the corresponding rate constants that accounts for the occurrence of the "slow regime" discussed above: for $t > 1$-1.5 ps, $E_{max}$ is already very close to $E_{max}^\infty$, whereas $W_G$ still changes, being 50-100 meV greater than $W_G^\infty$. Based on this observation, one can globally fit the time evolution for the entire red wing of the electron spectrum $S(E,t)$ for $t > 1$ ps using eq. (3) with $E_{max}(t) = E_{max}^\infty$ and

$$W_G(t) = W_G^\infty + \left(W_G^{max} - W_G^\infty\right)\exp(-t/\tau_G) \tag{4}$$

where $W_G^{max}$ is the Gaussian width extrapolated to $t = 0$ and $\tau_G$ is the time constant of the spectral narrowing. In Figs. 7 and 8 least-squares fits to the $S(E,t)$ kinetics for electron in light and heavy water, respectively, are shown for traces obtained between 0.85 and 1.3 µm. As seen from these plots, despite having only two adjustable parameters, $W_G^{max}$ and $\tau_G$, eqs. (3) and (4) give remarkably good approximation to these traces. The following

13.

estimates were obtained for these two parameters (95% confidence limits): $W_G^\infty$=1.18±0.04 eV and $\tau_G$=0.56±0.01ps (for H$_2$O) and $W_G^\infty$=1.2±0.03 eV and $\tau_G$=0.64±0.01 ps (for D$_2$O). Thus, in agreement with the analysis given in Fig. 6(b), the spectral narrowing in D$_2$O is slower than in H$_2$O, by ca. 15±5%.

Fig. 6(c) shows the evolution of the Lorentzian width $W_L$ of the absorption line as a function of the delay time $t_1$. After the first 1 ps following the photoionization, this width slowly increases from 0.4 to 0.5 eV, in approximately the same fashion for light and heavy water. When the integral under the spectrum profile is plotted vs. the delay time (Fig. 6(d)), it is almost constant with time (±10%), suggesting that in the "slow regime", the oscillator strength of the transition does not change with time: the narrowing on the red side is compensated by the broadening on the blue side. (However, it is difficult to establish this constancy accurately because the TA spectrum on the blue side is followed to $\lambda$=0.4 μm only).

We turn to the evolution of the spectra on the short time scale (0.5-1.2 ps). A curious feature of sub-picosecond TA spectra shown in Figs. 4(a) and 5(a) is an "isosbestic point" at 0.85 μm observed both in light and heavy water. [5,6] On closer examination, this feature does not constitute a true isosbestic point since the TA spectra obtained at different delay times do not pass through any particular point of the $e_{aq}^-$ spectrum within the confidence limits of our measurement. Some authors [e.g., 5,6,23,24] view this feature as evidence for the existence of two or more types of pre-solvated electron coexisting at short delay time; the observed TA spectra are then "decomposed" into separate contributions from the electron states with either time-dependent [e.g., 24] or time-independent [e.g., 5] spectra.

The largest impetus for the two-state model of the electron solvation was given by the original studies of Migus et al.[1,3] who observed a broad, short-lived component at 0.9-1.3 µm for the electron in $H_2O$ that decayed in the first picosecond after the multi-310 nm photon ionization. The overlap of this short-lived spectral component with the spectrum of the thermalized electron was thought to account for the "isosbestic point". Since the pump pulse in these pioneering studies was quite wide, this near-infrared feature was observed well within the duration of the photoexcitation pulse. Subsequent studies that used shorter pump pulses [e.g., 13,17] failed to reproduce this feature. For the electron in $D_2O$, the progression of the TA spectra shown in Fig. 5(a) has the same general appearance as that observed by Pépin et al.[6] following short-pulse multi- 600 nm photon ionization of $D_2O$. Their spectra, in turn, are qualitatively similar to the recent data of Unterreiner and co-workers[13] and Vilchiz et al.[17] for the electron in $H_2O$ who used short ultraviolet (UV) pulses generated by harmonic generation (on a Ti:sapphire laser system). Either the near-infrared feature observed by Migus et al.[1,3-5] is very short-lived (and, therefore, requires a relatively long excitation pulse to be observed against the background TA signal from the solvated electron) or, possibly, it is caused by nonlinear absorption of the probe light within the duration of the UV pulse (section 2).

The TA spectra obtained in this study within the duration of the excitation pulse also exhibit a diffuse absorption band in the near IR. In Fig. 6S, the spectra $S(E,t)$ for the electron in $D_2O$ shown for the 0-250 fs and 250-500 fs time windows. The visible data were excluded from the plot due to the strong contribution from nonlinear absorbance (the "spike" in Figs. 1(a) and 2(b)) that interferes with the measurement. While using the method of eq. (1) at these short delay times is not justified, the resulting spectra resemble the ones reported by Migus et al.[1] and Gauduel et al.[3,5] In Fig. 9, a near-infrared spectrum obtained in the point-to-point fashion within the last 10% of the duration of the 200 nm photoexcitation pulse is shown for the electron in $D_2O$. Once more, this spectrum is flat across the entire observation window; such a spectrum would be compatible with the existence of a short-lived ($e_{IR}^-$) species that strongly absorbs at $\lambda > 0.9$ μm. The transformation of this TA spectrum to that shown in Fig. 5(a) is very rapid, and we conclude that the coincidence of the excitation and probe pulses in time is needed to observe this diffuse TA band. Our data are insufficient to make a choice between the two possibilities discussed above (nonlinear absorbance vs. extremely short-lived species), and we leave the question open for further studies that use a shorter UV pulse.

Recently, Laenen et al. [16] obtained TA spectra in near- and mid- infrared after two 266 nm photon ionization of liquid $H_2O$ and $D_2O$. In $H_2O$, a broad absorption band from an intermediate (which the authors identify as "wet" electron) with a lifetime of 0.54 ps was observed at 1.5-1.7 μm. In $D_2O$, this band is red-shifted to 2 μm (still, the absorbance at 1.2 μm is ca 30% of the maximum). It is not clear from the published data how this feature evolves for $t > 0.5$ ps. Laenen et al. [16] give an estimate of $3 \times 10^4$ and $5 \times 10^4$ $M^{-1}$ $cm^{-1}$ for the molar absorptivity of the "wet" electron at 1.2 and 1.6 μm, respectively (in $D_2O$). Then, for these two probe wavelengths, a "wet" electron with a lifetime of 0.5-0.6 ps would yield a TA signal at $t = 0.5$ ps which comprises 30-80% of the absorption signal from a fully hydrated electron (having molar absorptivity of $2 \times 10^4$ $M^{-1}$ $cm^{-1}$ at the band maximum). [26,37] No such TA signal was observed in our study.

Instead, two features indicated by arrows in Fig. 4(a) were observed for the electron in light water: the 1.15 μm peak and the shoulder at 1.3-1.4 μm. Both of these features fully decay within the first picosecond after the photoionization, and their decay kinetics appear to be similar. Subtracting the 1.15 μm trace from the half-sum of the 1.1 and 1.2 μm traces, one obtains a kinetic profile that decays with a time constant of 0.2-0.3 ps, which is roughly the same time scale as that for the shift of the band maximum. The shoulder observed for $\lambda > 1.3$ μm might be the extension the 1.4-1.8 μm feature

16.

(from the putative "wet" electron) observed by Laenen et al.[16] It also looks similar to the 1.3 µm feature observed by Migus et al.[1] (a 0.24 ps life time was estimated for this feature therein).

We believe that these two features are not artifacts of the spectral reconstruction procedure described in section 2 (which is also suggested by the reasonable agreement with the previous studies); however, such a possibility cannot be fully excluded. Since both of these features are observed in the region where liquid $H_2O$ exhibits H-O vibration overtones, the concern is that the long-lived "absorbance" signal used for normalization of the TA kinetics is from the heat induced by the absorption of the UV light. The control experiments demonstrated that $e^-_{aq}$ scavenging by 1 M acid removes > 90-95% of the 1.0-1.3 µm signal, i.e., at most 5-10% of this long-lived TA signal could be from the thermal effect. This would be insufficient to account for the 1.15 µm feature which comprises ca. 20% of the TA signal at this wavelength. For the $\lambda > 1.3$ µm shoulder, such a test was difficult to conduct and we cannot discount the thermal effect entirely. It is, however, unlikely that the dynamics of such an effect would closely follow the dynamics of the 1.15 µm feature.

**4. Discussion.**

*4.1. The "slow regime".*

Why did the "slow regime" of the spectral evolution observed in section 3 escape the attention of previous workers? As explained in the Introduction, in most of the previous pump-probe studies, the spectral sampling was too sparse and the S/N ratio was inadequate to observe the subtle changes in the spectral shape. Furthermore, the best-quality spectral data were fit using the "continuous shift" model; for obvious reasons, such an analysis fails to identify the occurrence of the small-scale change in the spectral profile. Lastly, the "slow regime" has actually been observed previously, by Pépin et al.[6] These workers carried out a Gaussian-Lorentzian analysis similar to that outlined in section 3 for the electron generated by multiphoton ionization of liquid $D_2O$ ($\lambda$=0.5 to 1.4 µm). As seen from Figs. 3 and 4 in ref. 6, the Gaussian width systematically decreased with delay time, much like the same quantity in Fig. 6(b). To account for this apparent contradiction with the "continuous shift" model, Pépin et al. [6] postulated that the spectrum of the electron slides over a "supplementary component" from an unknown species having a life time of 2 ps that absorbs at 0.8-1.2 µm (see Fig. 6 therein). Subsequently, these observations were re-interpreted in terms of a "hybrid" model [23,24] in which two species, the "wet" electron and the solvated electron, both undergo continuous blue shift on the subpicosecond time scale; concurrently, the "wet" electron converts to the solvated electron. Analogies to the electron solvation in alcohols [25] were drawn to substantiate this model.

The "hybrid" model, much like the original "continuous shift" model, assumes that the spectral profile of the electron(s) does not change during the relaxation of the solvent (when the spectrum shifts) or, at the very least, this profile is a Gaussian-Lorentzian function. No justification has been given for either one of these two basic assumptions, and the "hybrid" model has as much fidelity as the "supplementary component" model suggested in ref. 6. Naturally, such elaborate models are nearly impossible to falsify because arbitrary properties can be postulated for hypothetical progenitors of various spectral components.

We believe that it would be more appropriate to interpret these observations at their face value, i.e., as evidence that the spectral profile of the ground-state electron does evolve with time and, at short delay times, deviates strongly from the Gaussian-Lorentzian form. As shown in section 3, the entire set of TA kinetics for the electron in $H_2O$ and $D_2O$ in the near infrared can be fit using the simplest spectrum-narrowing model with only two adjustable parameters.

Another concern is whether our observations can be explained using the "thermal jump" model of Madsen et al. [15] or the "cavity contraction" model of Unterreiner and co-workers [13] The answer is negative, because in their predictions, these two (seemingly different) models are almost indistinguishable from the "continuous shift" model. Within the framework of these models, the change in $E_{max}$ is interpreted either in terms of the decrease in the temperature of the water shell around the electron [15] or the mean square dispersion of its position with respect to the cavity center. [13] As for the latter model, the shape stability is the implicit assumption, i.e., the "cavity contraction" model is a reformulation of the "continuous shift" model. The temperature jump model, despite its different physical context, is also a variant of the "continuous shift" model: Since the width of the $e_{aq}^-$ spectrum changes with increasing temperature much slower than does $E_{max}$ (see Appendix A), simulations using this model yield spectral evolution which is very similar to that given by the "continuous shift" model. Furthermore, even a cursory analysis of the data in Figs. 6(a) and 6(b) points to a major inconsistency: while a temperature shift of 30 K would suffice to account for the position of the band maximum at $t = 0.5$ ps, a temperature shift of 200 K would be needed to account for the width of the electron spectrum (see Fig. 1S(b)). This conclusion, however, refers to the parameterization of the $e_{aq}^-$ spectra vs. temperature used by Madsen et al. [15] and this parameterization is not supported by more recent studies. In particular, Bartels et al. [37,38] observed substantial deviations of the spectral profile of hydrated electron in hot water from the Gaussian-Lorentzian curve given by eqs. (3) and (4) with the temperature-independent parameter $v = 2$. They demonstrated that in hot water, the spectral profile is given by a modified eq. (3) with $v$ that increases with the temperature and a temperature-independent width $W_L$. The resulting "hot-water" spectra become increasingly flatter at the top at higher temperature. [37,38] Given these recent developments, we re-analyzed the

19.

$S(\lambda,t)$ set in the spirit of the "temperature jump" model, using modified eq. (3) for constant $W_L$ and time-dependent parameters $E_{max}$, $W_G$, and $\nu$ (see Appendices A1. and A.3). As shown therein, the "temperature jump" model is not supported by our results, no matter which parameterization of the $e_{aq}^-$ spectrum is used.

The spectral narrowing described by eq. (3) complemented by eq. (4) (Figs. 7 and 8) naturally accounts for the major trend shown in Figs. 1, 2, and 3: the greater is the difference $E_{max}^\infty - E$ (> 0) for a given photon energy $E$, the faster is the decay of the electron absorbance at the corresponding wavelength. By increasing the bulk temperature or by addition of salt, it is possible to, respectively, red- or blue- shift the spectrum of hydrated electron (i.e., to decrease or increase $E_{max}^\infty$). [e.g., 39,40] Since the initial spectrum of pre-solvated electron does not change significantly as a function of temperature or salinity, the basic prediction of the spectrum-narrowing model (assuming weak temperature dependence for the relaxation time $\tau_G$ in eq. (4)) is that for a fixed probe wavelength on the red side of the $e_{aq}^-$ spectrum, the decay of the absorption is progressively slower with increasing temperature and progressively faster with increasing salinity. Both of these trends are consistent with the experimental observations of Unterreiner and co-workers [13] and Crowell and co-workers [33,41] (for the temperature effect) and Sauer et al. [39] (for the salinity effect).

### 4.2. Spectral evolution: possible mechanisms.

From the theoretical perspective, the chief appeal of the "continuous shift" model is a possibility to reduce the complex spectral evolution of the electron to a single time-dependent parameter $E_{max}(t)$ related to the average electron energy; the latter quantity may be obtained directly from mixed quantum-classical molecular dynamics (MD) simulations in which the excess electron is treated quantum-mechanically whereas the solvent molecules are treated classically.[42-49] If the "continuous shift" model were incorrect, such a simplified approach would be insufficient: time-resolved spectra of the electron should be simulated instead.[44,45] The latter is a much more complicated problem [e.g., 44-47] which requires the detailed knowledge of the ground and excited state, transition probabilities, etc. Even for fully hydrated, thermally relaxed electron, the origin of the basic features in its absorption spectrum is the subject of ongoing controversy and continuing research, despite the decades of modeling at all levels of theory, including, most recently, Car-Parrinello density functional MD modeling.[51]

The current consensus is that the bell-shaped band in the red is from three partially merged $s \rightarrow p$ subbands; these subbands correspond to the three orientations of the $p$ orbital with respect to the principal axes of the ellipsoidal solvation cavity.[43-45] While this simple picture is supported by MD[42-49] and other[51] calculations, hole burning absorption anisotropy experiments by Assel et al.[11] and, more recently, Cavanagh et al.[12] did not yield the anticipated polarization-dependent TA dynamics. Some anisotropy was observed in the experiments of Reid et al.,[52] however, even in that original study the depolarization dynamics were quite different from the theoretical predictions of Schwartz and Rossky.[45] A possible rationale for these discrepancies was proposed by Bratos and Leicknam[50] who postulated rapid (< 10 fs) internal relaxation in the $p$-state manifold following the photoexcitation of the $s$-state hydrated electron. Importantly, the rapid decay of the photon echo following the $s \rightarrow p$ photoexcitation observed by Wiersma and co-workers[10] cannot be considered as unequivocal evidence for fast dephasing in the $p$-state manifold since a rapid Stokes shift can also qualitatively account for their results.[49]

While a clear-cut demonstration of the substructure in the electron spectrum by means of ultrafast spectroscopy is lacking, indirect support for the $p$ state nondegeneracy is given by resonance Raman spectra of $e_{aq}^-$:[53-55] Tauber and Mathies[54] determined the depolarization ratio across the entire Raman spectrum of $e_{aq}^-$ in liquid $H_2O$ and concluded that this ratio (ca. 0.3-0.5) is incompatible with a resonant transition to a *single* nondegenerate state. Their analyses suggest that homogeneous broadening for $e_{aq}^-$ is at least 100 times greater than inhomogeneous broadening [53a,54] and the width of the $s \to p$ band on the red side is determined by vibrational progressions from the five strongest modes with wave numbers between 470 and 3100 cm$^{-1}$ (for $H_2O$). [53a] Thus, it appears that a full quantum mechanical treatment of water vibrations is needed to rationalize the shape of the absorption spectra.

It is apparent from these results that the absorption profile is controlled by many factors, and only advanced theory can account both for the spectrum of hydrated electron and the spectral evolution of its precursor in the course of solvation. As stated in the Introduction, given the previous record of *ad hoc* kinetic modeling, we do not feel compelled to provide yet another many-species reaction scheme "explaining" the solvation dynamics. Only a comprehensive, rigorous theory would be up to this task. Still, it is appropriate to speculate on the origin of the two novel features reported in this study, namely, (i) the spectral narrowing/broadening that occurs in the "slow regime" (for $t > 1$ ps) and (ii) the differences between the TA spectra for the excess electron in $H_2O$ and $D_2O$ (observed for $t < 1$ ps).

The characteristic spectral transformations observed in the "slow regime", *viz.* the narrowing of the band on the red side and broadening on the blue side, bear qualitative resemblance to the evolution of the absorption bands from electronically-excited molecules undergoing vibrational relaxation. Such a relaxation involves nonequilibrium populations of vibrational states which according to Tauber and Mathies [53,54] determine the shape and the width of the red side of the $e_{aq}^-$ spectrum. This vibrational relaxation readily explains the isotope dependence for the time constant of spectral narrowing. The

22.

time scale for the vibration relaxation in liquid water (ca. 0.75-1 ps for the O-H stretch in the room-temperature HOD:$D_2$O) [56] compares well with the observed time scale of spectrum narrowing in our experiment.

Another possible rationale is a time-dependent change in the relative positions and the weights of the $p$ subbands, although, the origin of the isotope effect is less clear in such a case. As follows from the results of Bartels et al. for hot and supercritical water, [37] the spectrum of $e^-_{aq}$ becomes flatter at the top with increasing water temperature. Similar trends were observed by Brodsky et al. [57] and Herrmann and Krebs [58] for methanol. The likely cause for this change is the increasing nonsphericity of the solvation cavity as it grows in size in the hot liquid. Consequently, the splitting between the three $p$ subbands increases and the electron spectrum becomes flatter at the top. [59] If the $p$ subbands do define the shape of the electron spectrum in the visible and near-infrared, as suggested by numerous MD calculations, the "slow regime" may be interpreted as gradual increase in the sphericity of the solvation cavity during its relaxation. That, in turn, suggests that the initial localization of the electron involves a contorted, highly anisotropic trap. Such an assertion is in agreement with the theoretical models of electron trapping in liquid water. [30] Indirect experimental support for this assertion is provided by the data on electron trapping in hexagonal ice: quasifree electrons were shown [19,60] to localize on (highly anisotropic) Bjerrum defects; [61] after this initial localization, the water structure around such a defect relaxes and a near-spherical solvation cavity gradually emerges on a picosecond time scale. [62] The resulting spectrum bears strong resemblance to the spectrum of hydrated electron and changes little between 4 and 270 K. [19,20]

It seems likely to these authors that the changing sphericity of the solvation cavity is the ultimate cause for the failure of the "continuous shift" and "temperature jump" models to capture the essentials of the spectral evolution on the subpicosecond timescale.

23.

In these models, the environment of the electron during its solvation is assumed to have the same degree of isotropy as fully relaxed, hydrated electron, either at the final or elevated temperature. Perhaps, the increase in the sphericity of the solvation cavity (both on sub-picosecond time scale and at later delay times) occurs in concert with the vibrational relaxation of solvent molecules forming the solvation cavity (see above); such a concerted mechanism would account for the isotope effect on the spectral narrowing.

On the subpicosecond time scale, the vibrations in water appear to have a strong effect on the spectral evolution. Indeed, for light water, new spectral features were observed in the near-infrared, where the water molecules have O-H stretch overtones. No such features were observed in $D_2O$ for which these overtones are red-shifted to $\lambda > 1.7$ μm. As explained in section 3, the thermal effect (a photoinduced change in the refraction index of water) does not appear to cause these features. Still, the vibrations of water molecules are implicated. Due to the extreme isotope selectivity, many potential candidates for the progenitors of these features, for instance, the IR-absorbing *p*-state electron examined theoretically by Schwartz and Rossky [45] and experimentally by Barbara and coworkers [7,9,52] and others, [11,12] can be excluded since the generation of such a species in the course of photoionization, while plausible, does not account for these specific features. We stress that our data (as well as the previous studies; see Kambhampati et al. [9] for more discussion) do not provide clear-cut evidence either against or for such an involvement, on a very short time scale (< 300 fs). The same refers to the involvement of the tentative "wet" electron [6,11,16,22-25] and liquid-water analog of the IR-absorbing electron in low-temperature ice [19-21] postulated by others.

We believe that the 1.15 and 1.3-1.4 μm features for the excess electron in $H_2O$ emerge due to collective excitation of the electronic and vibrational modes of the pre-solvated electron by the probe light, when the corresponding energies are close. It is

24.

possible that the same features are present in the TA spectra of $s \rightarrow p$ excited hydrated electrons; unfortunately, no spectra in the near- and mid-infrared are currently available (only 0.55-1.05 μm spectra have been reported). [7-9,11] In the photon echo and transient grating experiments of Wiersma and co-workers, a librational motion of water molecules (ca. 850 cm$^{-1}$) in the relaxation dynamics of electrons following $s \rightarrow p$ photoexcitation was observed. Resonance Raman spectra of the hydrated electron indicate strong coupling of the $s \rightarrow p$ electronic transition to the O-H stretch and bend modes. [53-55] Such a coupling would be expected on the theoretical grounds since the O-H groups with protons pointing out towards the cavity center comprise the core of the solvated electron. [42-47] Vibronic transitions in the near-infrared would be another manifestation of this strong coupling. Specifically, we suggest that O-H vibrations can be excited when the energy of the electronic transition is close to that of the O-H stretch overtone. To our knowledge, such a possibility has not been addressed theoretically, while it might be suggested by the results of the present work. In the currently popular mixed quantum-classical MD [43-47] and path integral [42] models of excess electrons in liquids, the solvent is treated classically and such phenomena as vibronic coupling and vibrational relaxation for hydrated electron cannot be addressed. The results discussed above suggest that these phenomena might play important roles in the electron solvation dynamics; hence, more advanced models might be needed.

## 5. Conclusion.

The evolution of the TA spectra in the visible and near-infrared, for the pre-solvated electron, following biphotonic excitation of room-temperature light and heavy water, has been studied. Two regimes of the spectral evolution were observed. In both of these regimes, the spectral profile changes with delay time; the "continuous blue shift" [e.g., 6, 22] and "temperature jump" models [14] that assume the constancy of this spectral profile (as the electron spectrum shifts to the blue) are not supported by our data. On the sub-

25.

picosecond time scale, the band maximum systematically shifts to higher photon energy and the spectrum narrows (Figs. 4(a), 5(a), and 5S). At later delay times (> 1ps), the position of the band maximum is "locked", but the spectral profile continues to change, by narrowing on the red side and broadening on the blue side (Figs. 4(b) and 5(b)). The oscillator strength of the transition is constant within 10% during this relatively slow, small-scale spectral transformation. The spectral narrowing can be accounted for by a two-parameter model in which the width of the Gaussian half-line exponentially decreases with the delay time. The time constant of this narrowing is ca. 0.56 ps for $H_2O$ and 0.64 ps for $D_2O$, respectively.

In the first picosecond after photoionization, two new features (the 1.15 μm band and 1.4 μm shoulder) were observed in the near-infrared spectra for the electron in $H_2O$; these two features were not observed for the electron in $D_2O$. These features are observed in the same region were O-H overtones appear in the spectrum of liquid water. While it cannot be entirely excluded that these two features are artifacts of the data analysis, it seems more likely that these bands are genuine and originate through the vibronic transitions of pre-solvated electron. The isotope-dependent narrowing/broadening of the absorption band of the electron for $t >1$ ps can be interpreted as evidence for the occurrence of vibrational relaxation in the water molecules lining the solvation cavity. In both of these regimes, continuous decrease in the size and increase in the sphericity of the solvation cavity causes a time-dependent blue shift and decrease in the splitting between the three $p$ subbands of the electron. These two trends would qualitatively account for the observed spectral evolution.

Since the current MD theories of hydrated electron [42-49] do not include the quantum degrees of freedom for vibrations in water molecules, the vibrational relaxation and the vibronic coupling hinted at by our results presently cannot be modelled. While there are models which treat the vibrations quantum-mechanically, albeit in a greatly simplistified fashion, [57,63] these models cannot address the solvation dynamics. Further development of the theory would be necessary for a self-consistent explanation of the spectral evolution during the solvation and thermalization of the electron in liquid water.

**6. Acknowledgement.**

We thank Profs. B. J. Schwartz and P. F. Barbara and Drs. S. Pommeret, C. D. Jonah, and D. M. Bartels for many useful discussions. We thank Dr. D. M. Bartels for the permission to reproduce his unpublished data. The research at the ANL was supported by the Office of Science, Division of Chemical Sciences, US-DOE under contract number W-31-109-ENG-38.

*Supporting Information Available:* (1.) A PDF file containing (a) Appendix A. The absorption spectrum of hydrated electron, (b) Appendix B. Instruction for the retrieval of kinetic data from the supplied ascii file, (c) Figs. 1S to 7S with captions. (2.) A 288 Kb ascii file named "H2O_D2O_traces.txt" containing digitized kinetic traces $S(\lambda,t)$ for the electron in $H_2O$ and $D_2O$. This material is available free of charge via the Internet at http://pubs.acs.org.

**References.**

**Figure captions.**

**Fig. 1**

A family of kinetic traces $S(\lambda,t)$ (section 2 and eq. (1)) for electron generated by bi- 200 nm photonic ionization of the room-temperature light water. The magnitude of the hydrated electron spectrum at the maximum is taken as unity. The probe wavelength $\lambda$ (500 to 1350 nm) is given in the color scales to the right of the plots. The vertical bars are 95% confidence limits. The "spike" near the kinetic origin in (a) is from nonlinear absorbance due to simultaneous absorption of 200 nm and probe photons. The sigmoid curve in (b) is the integral of the response function of the system (which is shown by a shaded curve in (c)).

**Fig. 2**

Same as Fig. 1, for heavy water. Only 500 to 1050 nm data are shown.

**Fig. 3**

Same as Fig. 1, for heavy water. Only 1100 to 1700 nm data are shown. In (b), the vertical scale is expanded to illustrate nonzero plateau attained at $t > 3$ ps.

**Fig. 4**

TA spectra $S(\lambda,t)$ for electron in light water at different intervals $(t_1, t_1 + \Delta t)$ of delay time. The delay times $t_1$ are given in the plots. For (a) $\Delta t = 100$ fs, for (b) - 200 fs. The maximum TA signal from the hydrated electron $S(\lambda) = S(\lambda, t = \infty)$ is taken as unity. The solid lines in (b) are Gaussian-Lorentzian functions obtained by least-squares fits, as explained in the text (note the logarithmic vertical scale). The vertical bars are 95% confidence limits. The bold solid line is the normalized spectrum ($S(\lambda)$) of fully thermalized hydrated electron. The vertical arrows indicate previously unobserved features in the near-infrared.

**Fig. 5**

Same as Fig. 4, for the electron in liquid $D_2O$.

**Fig. 6**

Gaussian-Lorentzian analysis of linearly interpolated $S(\lambda,t)$ data for the electron in light *(open squares)* and heavy *(open circles)* water. Vertical bars are 95% confidence limits determined by least-squares optimization. The solid lines in (a) and (b) are exponential fits. (a) The position of the spectrum maximum, (b) Gaussian width, (c) Lorentzian width, and (d) the integral under the spectrum plotted as a function of the delay time $t_1$ for $\Delta t = 30$ fs.

**Fig. 7**

"Global" least squares analysis of $S(\lambda,t)$ kinetics using the spectrum-narrowing model (eqs. (3) and (4)) for the electron in light water. The solid lines are simulated kinetics for, *from top to bottom*: $\lambda = 0.8$, 0.85, 0.9, 0.95, 1, 1.05, 1.1, 1.2, 1.25, and 1.3 µm. The vertical bars are 95% confidence limits.

**Fig. 8**

Same as Fig. 7, for the electron in heavy water. The probe wavelengths are 0.85, 0.9, 0.95, 1, 1.05, 1.1, 1.2, 1.25, and 1.3 µm.

**Fig. 9**

*Open circles:* TA signal for the electron in $D_2O$ vs. the wavelength $\lambda$ of the probe light ($\lambda > 1.1$ µm). The delay time between the 200 nm pump and the probe pulses was fixed (this delay time corresponded to the last 10% of the pump pulse duration). The solid line is a guide to the eye.

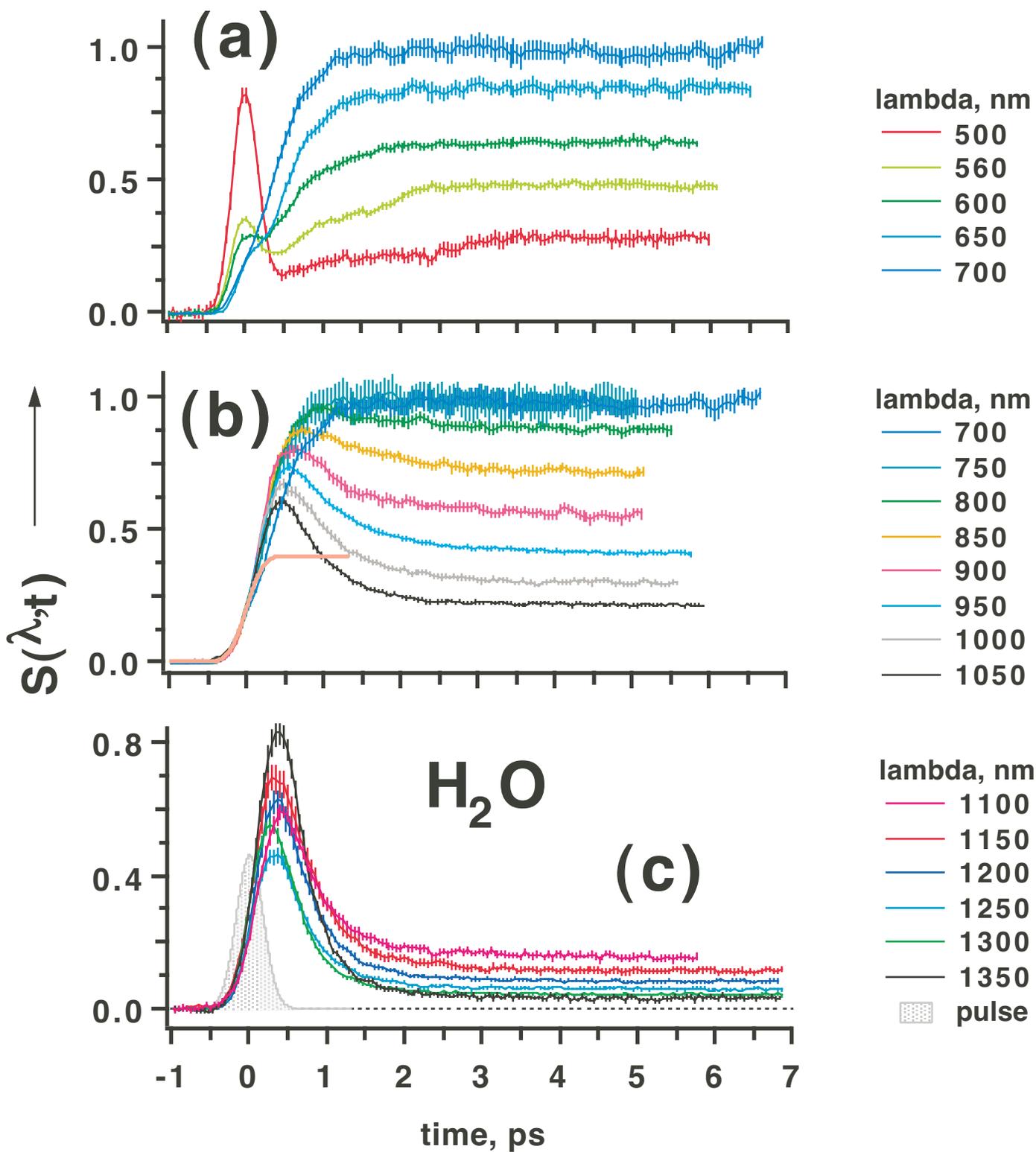

Lian et al., Figure 1

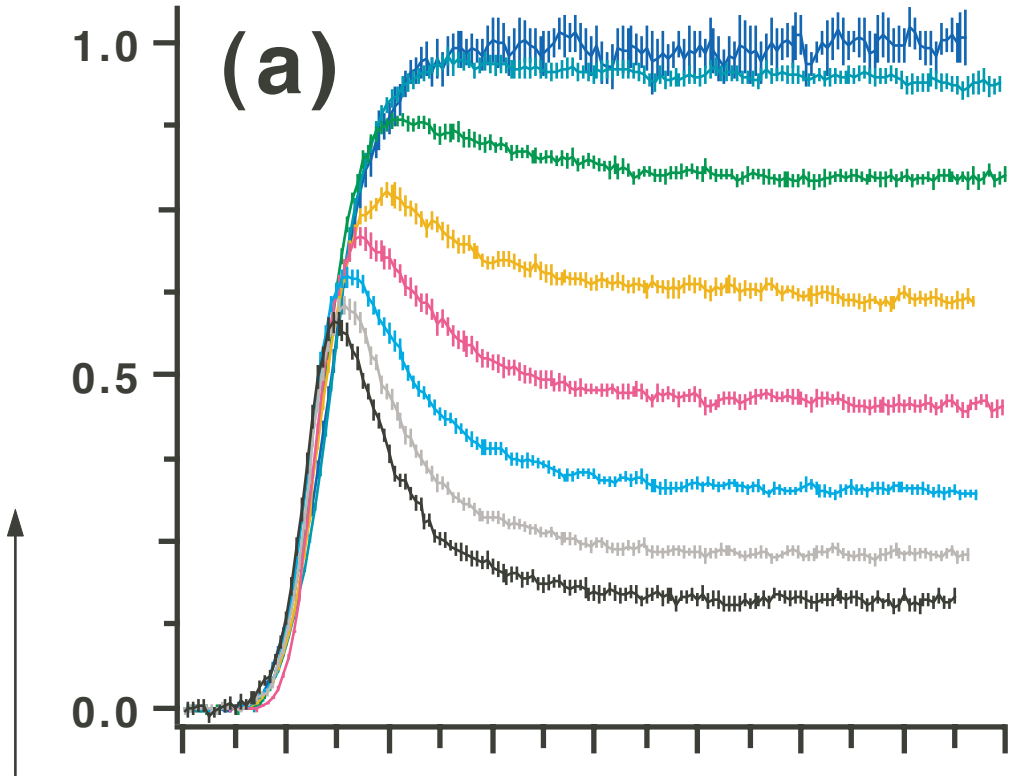
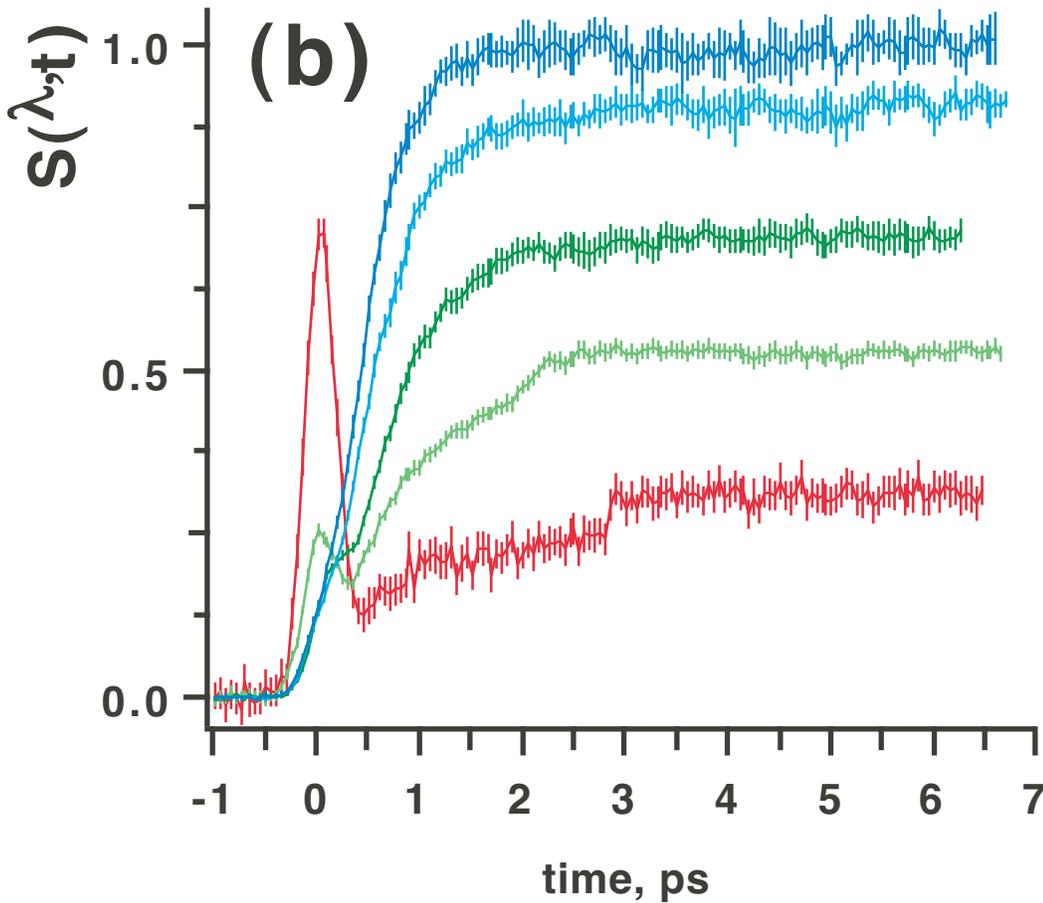

Lian et al., Figure 2

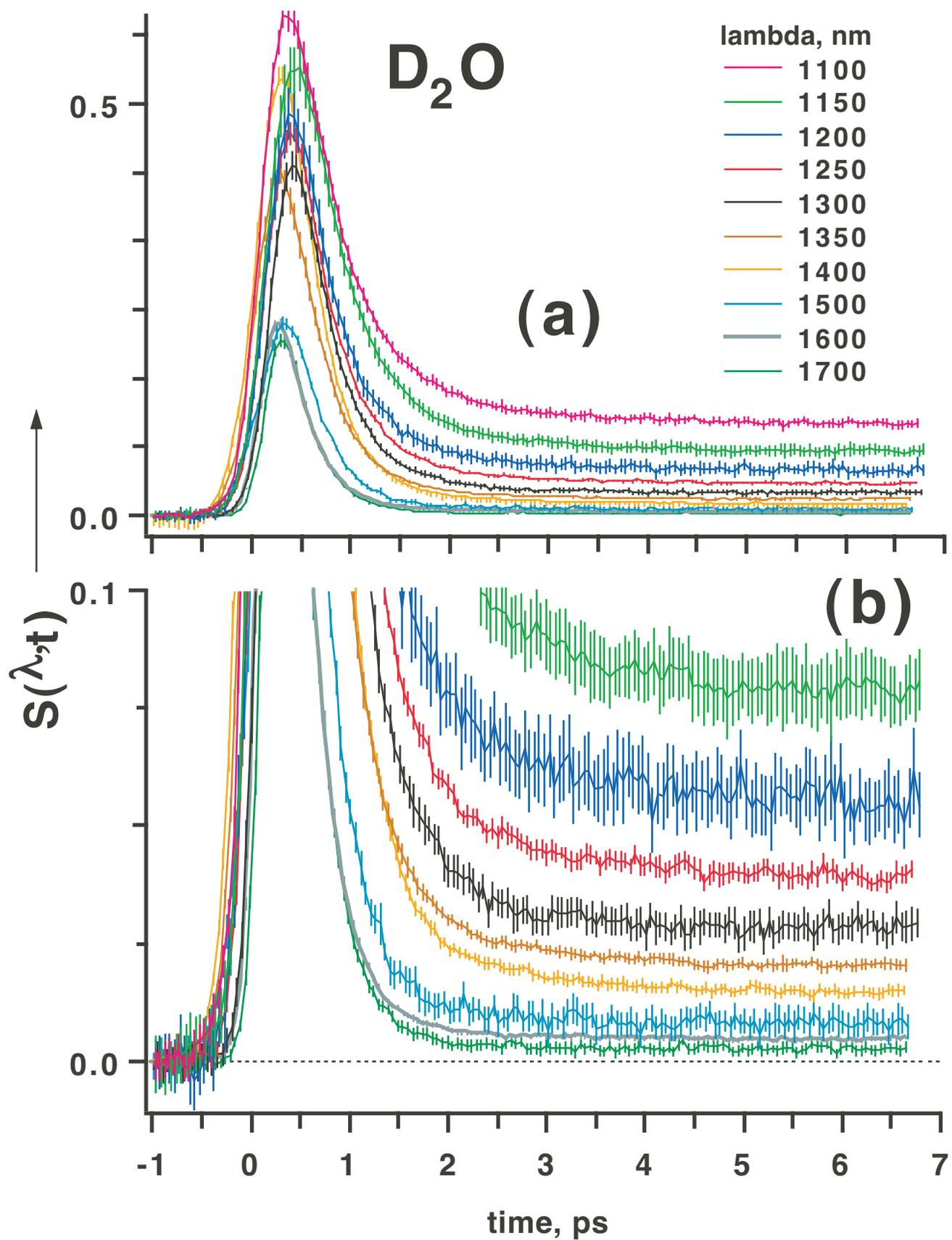

Lian et al., Figure 3

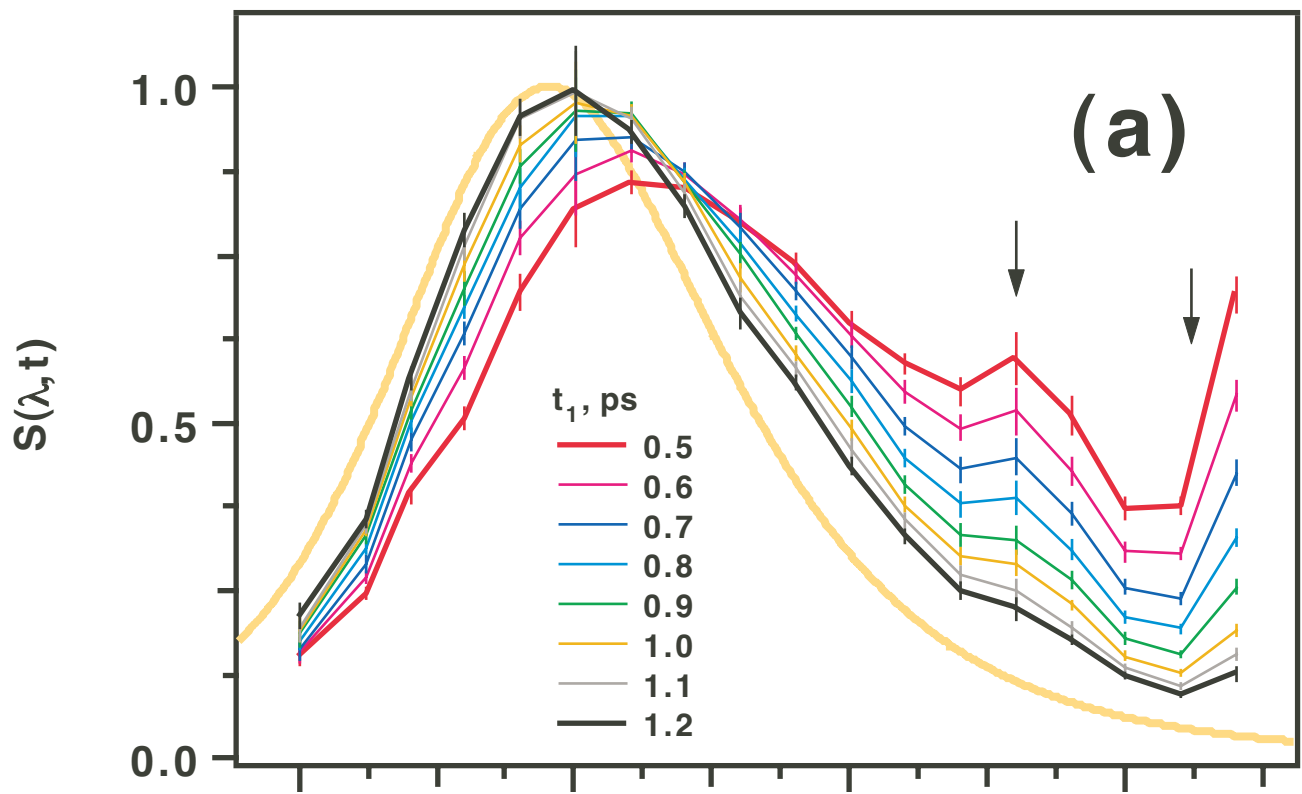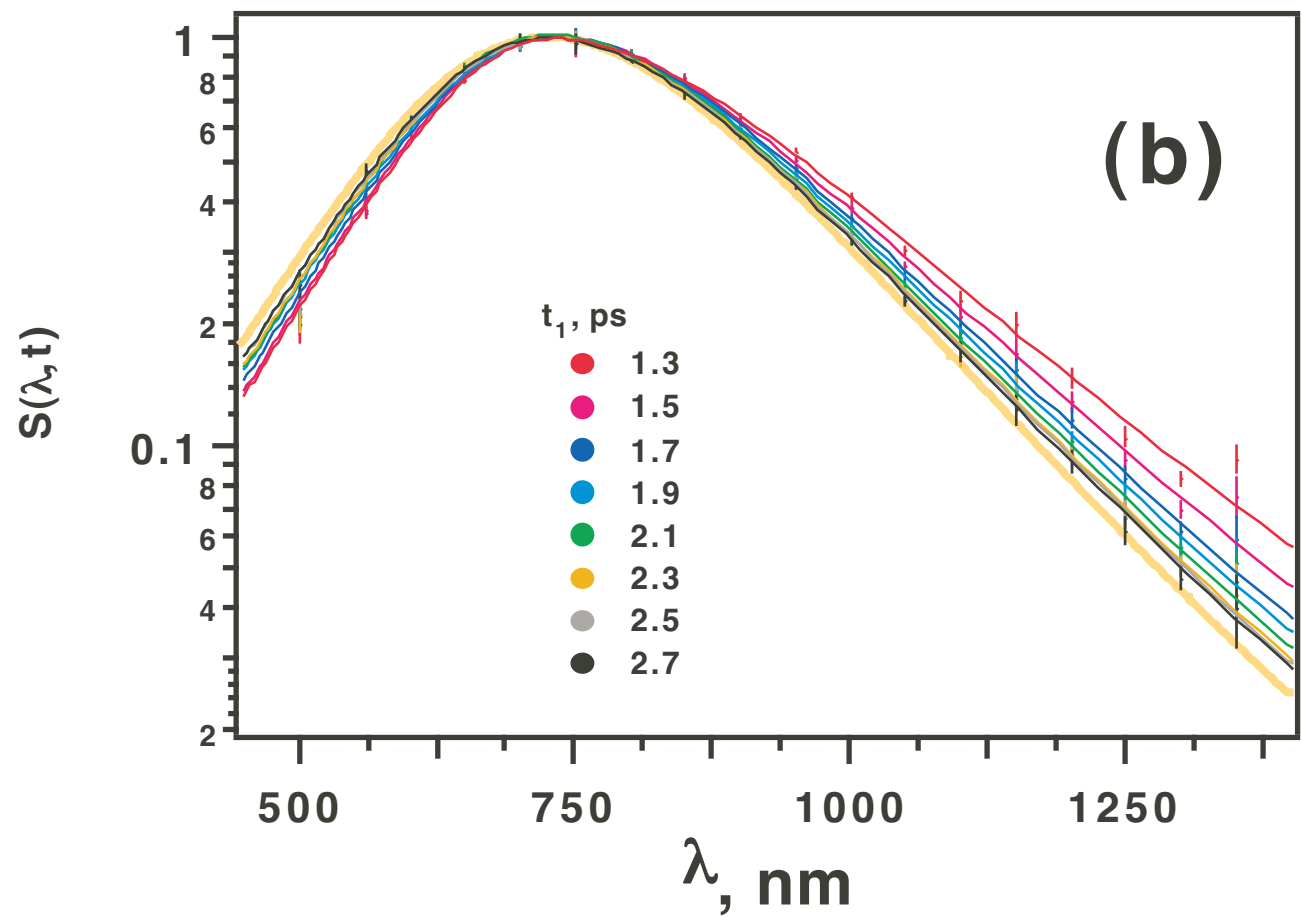

Lian et al., Figure 4

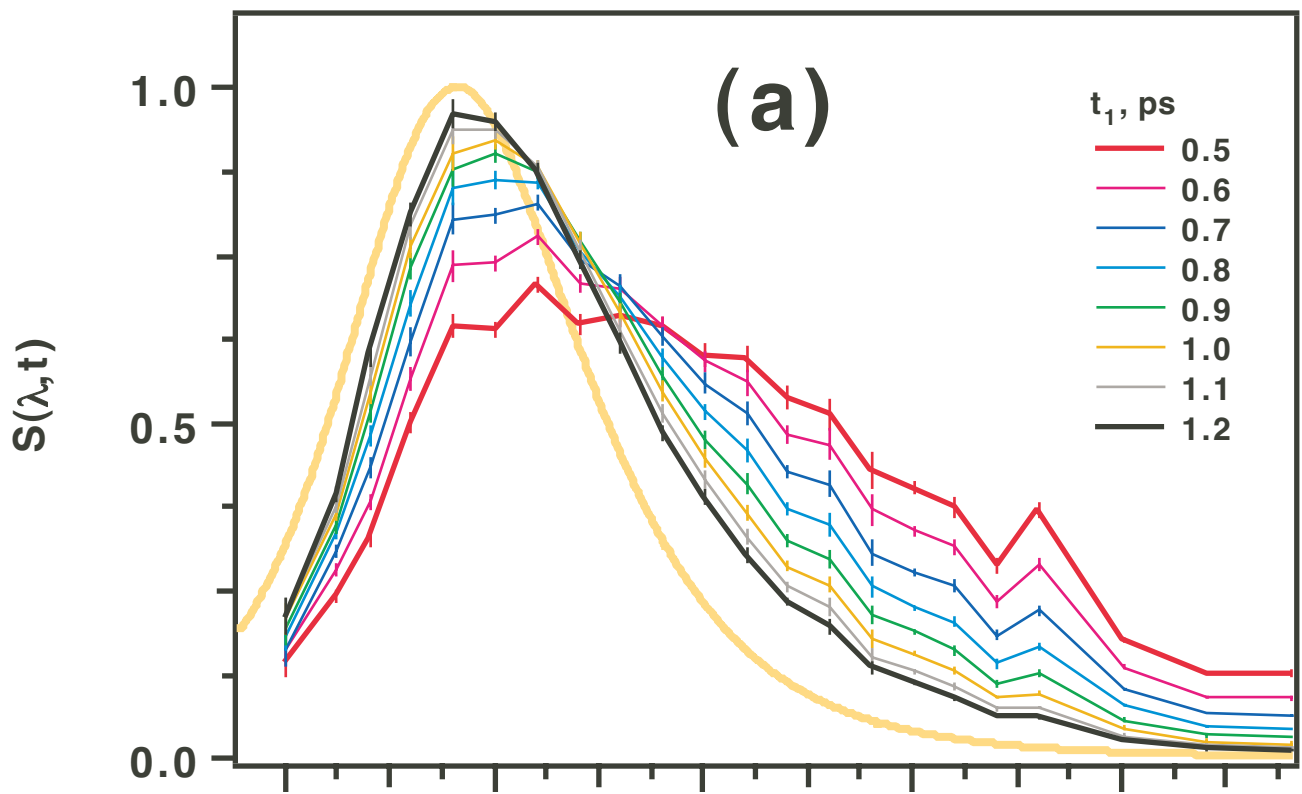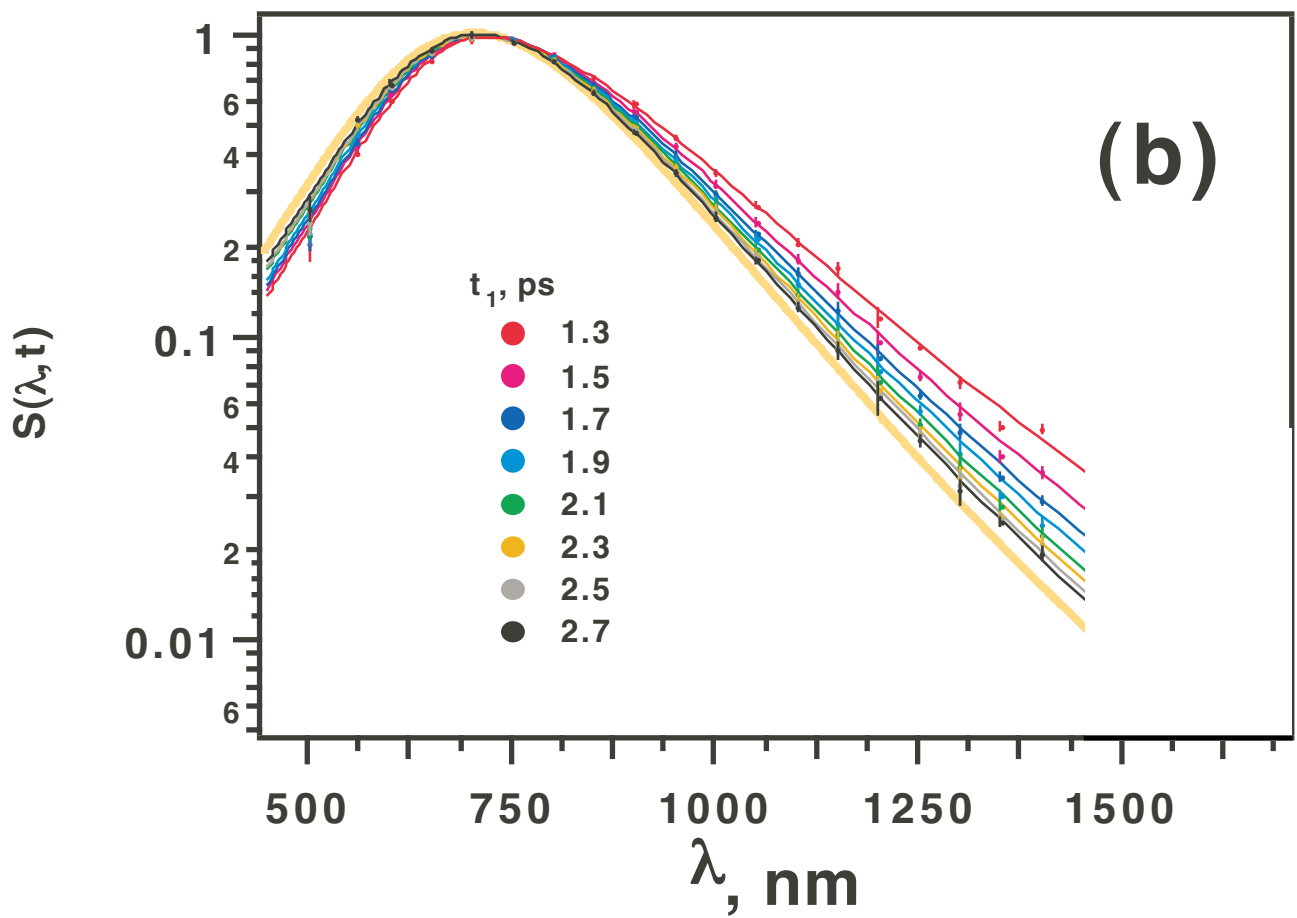

Lian et al., Figure 5

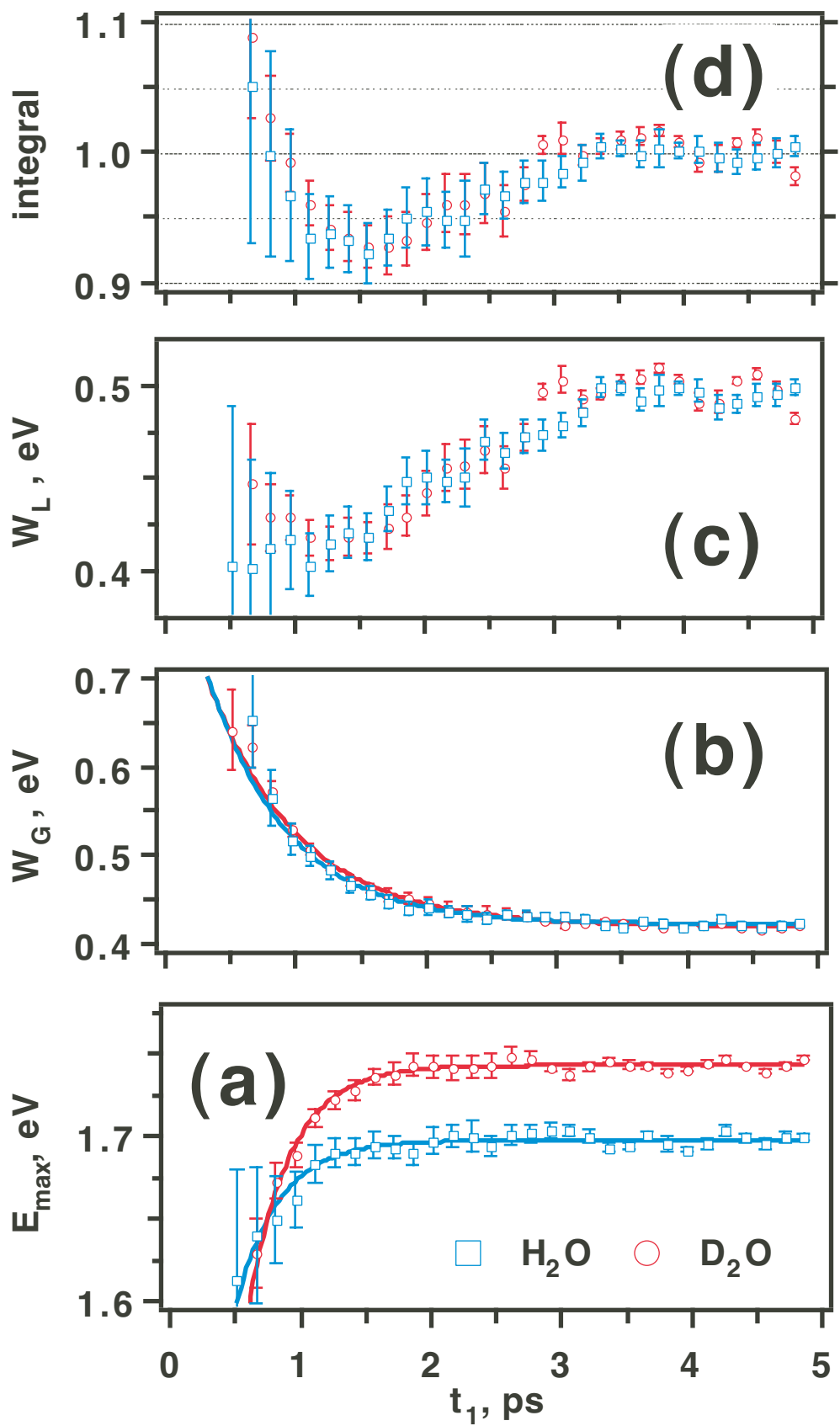

Lian et al., Figure 6

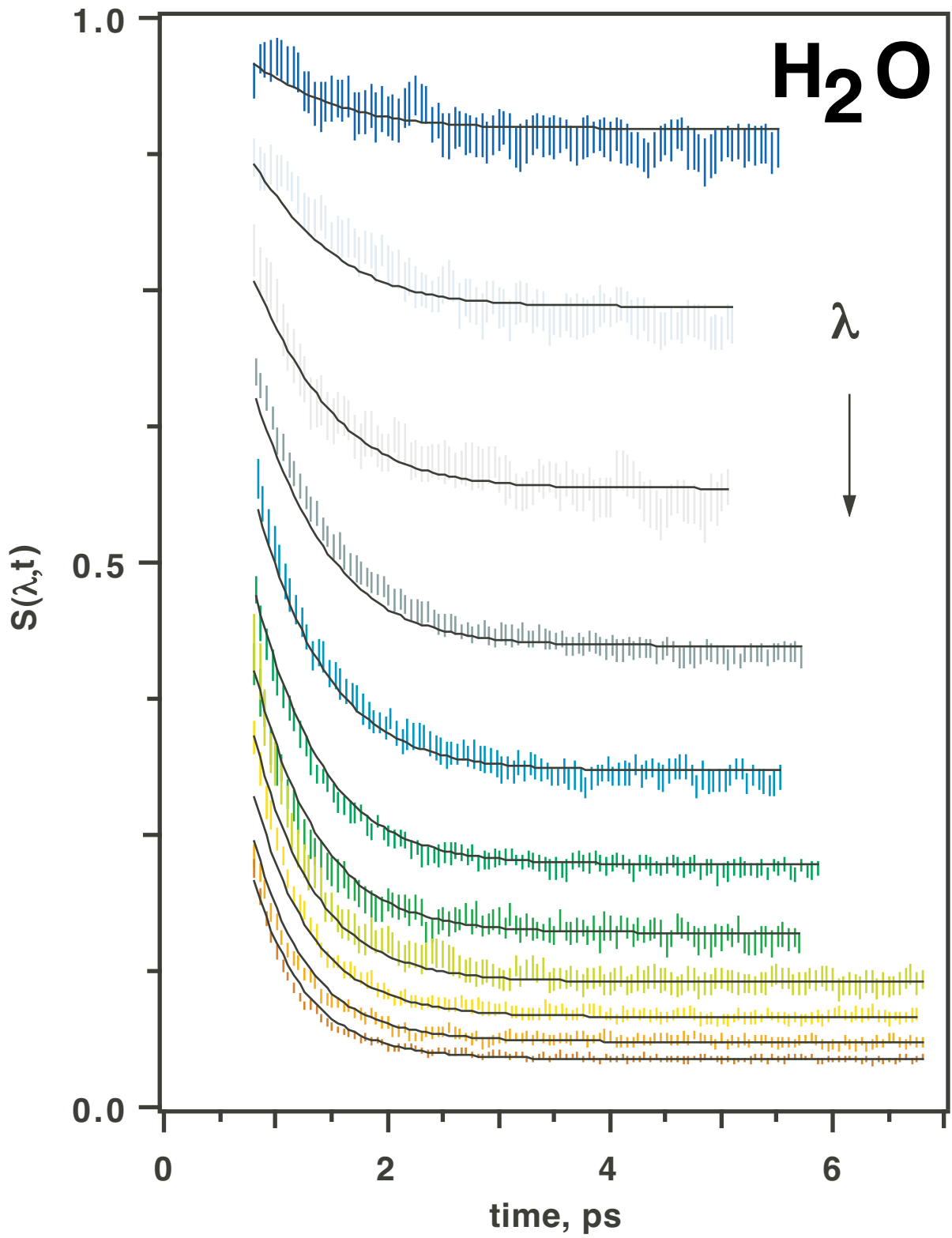

Lian et al., Figure 7

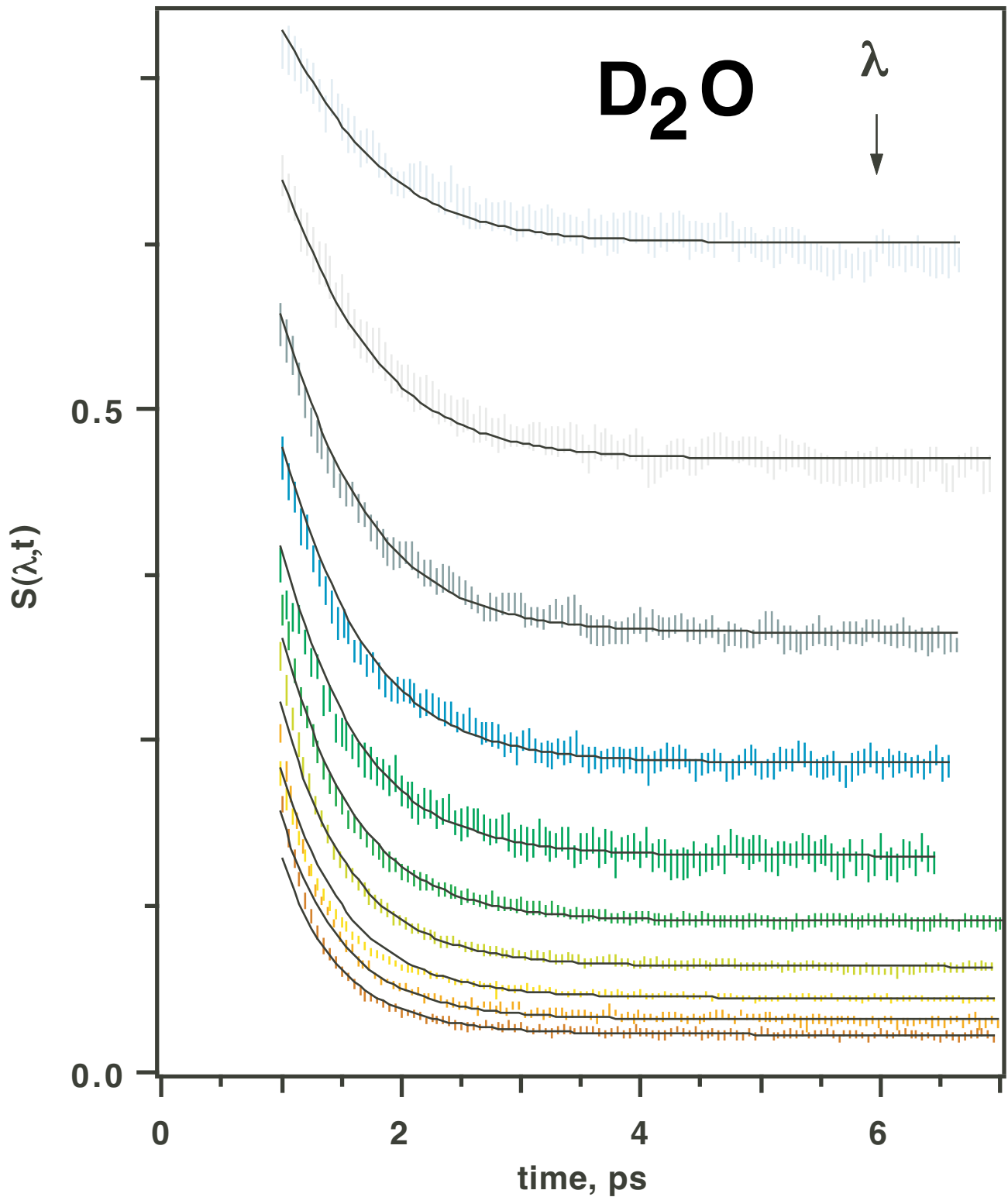

Lian et al., Figure 8

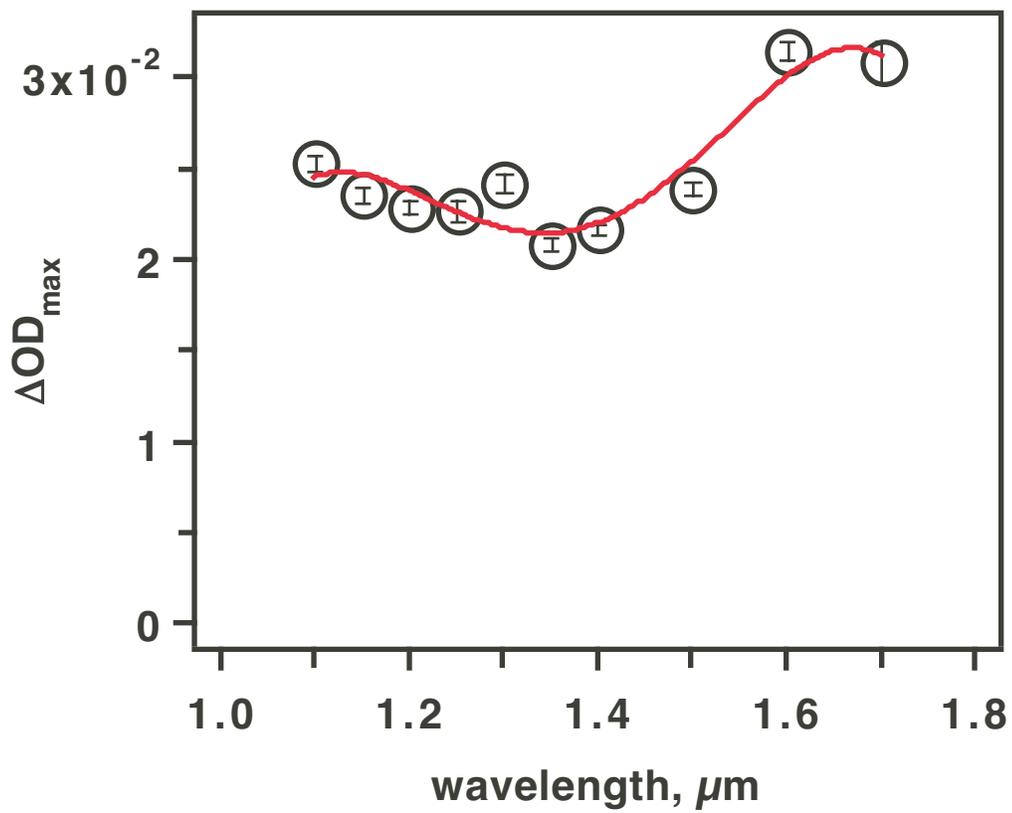

Lian et al., Figure 9

**SUPPLEMENTARY MATERIAL**                                        JP0000000

*Journal of Physical Chemistry A, Received ****

**Supporting Information.**

**Solvation and thermalization of electrons generated by above the gap (12.4 eV) two-photon ionization of liquid $H_2O$ and $D_2O$.**

Rui Lian, Robert A. Crowell, and Ilya A. Shkrob.

*Chemistry Division, Argonne National Laboratory, Argonne, IL 60439*

**Appendix A. The absorption spectrum of hydrated electron.**

*A.1. Background.*

Although many absorption spectra of hydrated electron in $H_2O$ and $D_2O$ have been reported in the literature, there are very few such spectra that exhibit acceptable S/N ratio in the near IR. Good quality data in this spectral region are needed for the reconstruction of the time-dependent TA spectra of pre-solvated electron using the method discussed in section 2. The prime concern is the accuracy of the Gaussian fit to the low-energy wing of the $e^-_{aq}$ spectrum.

Recently, Bartels and coworkers [37,38] obtained a large set of spectral data for radiolytically-generated $e^-_{aq}$ at 25-300 °C and fitted this set globally to eqs. (2) and (3) using the temperature-dependent exponent

$$v = 1.9488 + 0.0012895t. \tag{A1}$$

where $t$ is the temperature in °C. For $H_2O$, they obtained

$$E_{max}(eV) = 1.79 - 0.0025407t + 2.5384 \times 10^{-17}t^5, \tag{A2}$$

$$W_G(eV)/\sqrt{\ln 2} = 0.343 + 0.000290411t + 6.613 \times 10^{-6}t^2 \\ - 4.10413 \times 10^{-8}t^3 + 5.33721 \times 10^{-11}t^4, \tag{A3}$$

and $W_L$=0.51 eV. For $D_2O$, the width $W_L$=0.49 eV and

$$E_{max}(eV) = 1.84 - 0.00268t + 4.13 \times 10^{-17}t^6, \tag{A4}$$

$$W_G(eV)/\sqrt{\ln 2} = 0.337 + 5.5 \times 10^{-4}t + 2 \times 10^{-8}t^3 \\ - 1.9 \times 10^{-10}t^4 - 3.55 \times 10^{-13}t^5, \tag{A5}$$

To obtain eqs. (A1) to (A5), the $e_{aq}^-$ spectra acquired in steps of 50 nm between 0.3 and 1.7 μm were fit using eqs. (2) and (3) with the floating parameters $v$, $W_G$ and $E_{max}$ using the least squares method; the optimum parameters were then polynominally interpolated.

Figs. 1S(a) and 1S(b) show the temperature dependencies of the parameters $v$, $W_G$, and $E_{max}$ given by eqs. (A1) to (A5). As seen from the $v(t)$ plot given in Fig. 1S(a), the high-energy wing of the electron spectrum ($E > E_{max}$) is Lorentzian ($v \approx 2$) only near room temperature; e.g., at 300 °C, $v \approx 2.35$ and the spectra are much flatter at the top than the room-temperature spectra, for which $v \approx 1.95$. In the previous studies, (e.g., ref. 26) the shape of the spectrum for $E > E_{max}$ was assumed to be Lorentzian with a temperature-invariant exponent $v = 2$; this assumption is no longer tenable. The variation in the Gaussian width $W_G$ of the half-spectra with increasing temperature is nonnegligible (ca. 20% over the temperature interval of 200 °C), although it is much less than the decrease in $E_{max}$ (ca. 50% over the temperature interval of 300 °C). It is this relative constancy of the spectral profile vs. the temperature which provides the rationale for the "continuous shift" and "temperature jump" models discussed in sections 1 and 4.2.

The blue side of the electron spectrum ($E > E_{max}$) is less perfectly approximated by eq. (2) than the red side ($E < E_{max}$) is approximated by eq. (3), whether a fixed ($v = 2$) or variable parameter $v$ is used (Fig. 1S(a)). While on the red side the absolute deviation of the data points from the Gaussian shape eq. (3) is less than 2-3 %, the deviation on the blue side is typically 5-10 %, depending on the wavelength of the analyzing light. The further into the UV, the worse is the overall fit quality. For this reason, the global least-squares fit is a trade-off: the better is the fit quality for the extension of the spectrum to the near-IR, the worse is the fit quality for the extension of this spectrum to the blue, and vice versa. This leads to uncertainty in the position of the absorption maximum $E_{max}$, which is reflected in the scatter of the reported values. Since the electron spectrum is flat at the top, the location of this maximum depends on the exact functional form of the prescribed profile of the spectrum. Inasmuch as this location depends on how well the Lorentzian curve approximates the data points for $E > E_{max}$, the position of this maximum depends on the extent of the spectrum to the blue. As the quality of the spectral reconstruction given by eq. (1) in section 2 crucially depends on the accuracy of the $e_{aq}^-$ spectrum in the near IR, parameterizations were obtained that are most accurate in that particular spectral region. This can only be achieved by truncating the $e_{aq}^-$ spectrum at 400 nm.

### A.2. Electron spectrum from flash photolysis data.

To obtain a good-quality spectrum of $e_{aq}^-$, hydrated electrons were generated by 248 nm (one-) photon excitation of 330 μm ferrocyanide (hexacyanoferrate (II)) and 4.8 mM sulfite using 15 ns FWHM, 10 mJ pulses from an ArF excimer laser (Lambda Physik model LPX120i) in a 1.36 mm optical path cell with suprasil windows. These two aqueous anions photoreact according to

$$Fe(CN)_6^{4-} \xrightarrow{h\nu} Fe(CN)_6^{3-} + e_{aq}^- \tag{A6}$$

and

$$SO_3^{2-} \xrightarrow{h\nu} SO_3^- + e_{aq}^-. \tag{A7}$$

The anion absorptivities and the quantum yields for electron photodetachment are 4270 M$^{-1}$ cm$^{-1}$ and 0.674 for ferrocyanide and 50 M$^{-1}$ cm$^{-1}$ and 0.108 for sulfite, respectively.[36] The advantage of using ferrocyanide is the high electron yield; the disadvantage is that *ferri*cyanide anion (hexacyanoferrate (III)) generated in rxn. (A6) absorbs analyzing light below 450 nm. Thus, to obtain the electron spectrum in the blue, electron photodetachment from sulfite should be used (the $SO_3^-$ radical generated in rxn. (A9) does not absorb above 400 nm).

The 248 nm beam normal to the window uniformly illuminated a 3 mm x 6 mm aperture placed onto the front window of the photocell. The analyzing light from a pulsed 75 W Xe lamp was crossed at 30° with this beam. After traversing the sample, the analyzing light passed through an appropriate cutoff filter and was focussed on the photodiode using a 7 cm focal length achromat. The wavelength of the analyzing light was selected using the same narrowband interference filters used in our pump-probe kinetic studies. The signal from the fast Si and Ge photodiodes (of the same type as in section 2) loaded into 50 Ω was amplified using a 1.2 GHz video amplifier (Comlinear model CLC449) and sampled by a digital signal analyzer (Tektronix model DSA601). The response time of this detection system was 3 ns. 0.2-1 L of aqueous N$_2$-saturated solution of K$_4$Fe(CN)$_6$ or Na$_2$SO$_4$ (99.99+% ultrapure grade, Aldrich) was circulated through the cell using a peristaltic pump. The typical flow rate was 2-3 cm$^3$/min; the repetition rate of the laser was 1 Hz. All measurements were carried out at 24 °C. Each kinetic trace was an average of 20-30 laser shots. The time profile of these kinetic traces (for wavelengths $\lambda$ where only electron absorbed) was independent of $\lambda$; no effect of continuous exposure of the sample to the UV light was observed. The half time of the electron decay was 400 ns for ferrocyanide and 5 μs for sulfite, respectively. The life time of $e_{aq}^-$ was controlled by recombination in the bulk; for sulfite, a slow scavenging reaction with an impurity (likely, traces of oxygen) was also significant. The time-dependent TA signal was integrated between 50 and 120 ns and these integrals (normalized by the integral at 700 nm) were plotted vs. $\lambda$ and fit using eqs. (2) and (3) with $\nu = 2$.

The composite spectrum of $e_{aq}^-$ in H$_2$O is shown in Fig. 2S. For $\lambda$<500 nm, the data for sulfite and ferrocyanide photolysis are both shown; at higher photon energy, only sulfite data are shown. The optimum fit parameters obtained for this data set are $E_{max}$=1.699±0.005 eV, $W_G$=0.422±0.005 eV, and $W_L$=0.492±0.007 eV (the standard deviations are given). The data for D$_2$O (only low-energy "tail" of the spectrum is shown in Fig. 3S) can be fit using the same $W_L$ and $W_G$ and $E_{max}$=1.749 eV (which is ca. 3% higher than the light water estimate). As shown in Figs. 2S and 3S, the residuals are less than 0.02. For $\lambda$<500 nm, our data for $e_{aq}^-$ in H$_2$O compare well with the data given by Jou and Freeman [26] (see Table 1S), although the fit parameters are different. Ironically, using the "optimum" parameters given by Jou and Freeman [26] makes the fit quality in the near-IR worse, because their parameters were, apparently, optimized to provide the best fit in the visible and the UV. The same applies to the parameterization of the $e_{aq}^-$ spectrum by Bartels and coworkers: [37,38] As shown in Table 1S and Fig. 2S, the $S(E)$ curve calculated using the spectral parameters given by eqs. (A1) to (A5) fits well the experimental data on the blue side of the spectrum but much worse on the red side of the same spectrum. On the red side, $S(E)$ is systematically underestimated by as much as 5%.

Since the reconstruction of the absorption spectrum of pre-thermalized electron by the method discussed in section 2 critically depends on the quality of the $e_{aq}^-$ spectrum in the near IR (where $e_{aq}^-$ absorbs poorly), the parameterization given above is preferable to that given by others. Given that in the near IR, the ratio of the TA signal in the "spike" to the TA signal from the thermalized electron at $t$=5 ps can be as large as 20:1 (Fig. 3), even a small error (1-2%) in the $e_{aq}^-$ spectrum $S(\lambda)$ used for the normalization translates into a 10-50% error in the amplitude of $S(\lambda,t)$ at short delay time. These large error bars undermine the confidence in the spectral data obtained using the normalization method on a very short time scale. Other reasons for treating such data with caution, *viz.* the thermal effects and the wavelength dependence of the probe pulse characteristics, are briefly discussed in sections 2 and 3 in the text.

While there are several alternatives to the (obviously, imperfect) normalization method used in this work, these methods proved to be equally problematic for obtaining a good-quality sub-picosecond TA spectrum over a sufficiently wide wavelength range. Scanning the wavelength of the probe light for a fixed delay time, by using a set of interference filters or a monochromator (as done to generate the spectrum shown in Fig. 9), does not correct for the wavelength-dependent change in the width (i.e., the chirp) in the probe pulse and the group velocity mismatch (see ref. 17 for discussion). Obtaining two-dimensional TA data using diode array or CCD detectors makes it possible to track these changes better, but the dynamic range of these detectors is small, and it is impossible to cover a large section of the $e_{aq}^-$ spectrum in the near-IR, where the intensity of transmitted light decreases precipitously with increasing wavelength. Using optical parametric amplifiers instead of the white-light supercontinuum generation leads to the same difficulties when $\lambda$ is changed over a sufficiently wide range, especially in the near- and mid- IR (see, e.g., ref. 16). Since the pulse characteristics continuously change with the probe wavelength, it is impossible to interpret the data without prescribing an *ad hoc* kinetic model, and this makes the resulting TA spectra dependent on the details of such a model. Our goal was to eliminate such a dependency, hence our reliance on the (imperfect) normalization method.

***A.3. "Temperature jump" model.***

Given the developments discussed in section A.1, we re-analyzed our $S(\lambda,t)$ set in the spirit of the "temperature jump" model of Madsen et al., [15] using modified eq. (3) for constant $W_L$ and time-dependent parameters $E_{max}$, $W_G$, and $v$ (section A.1). Figs. 6S and 7S show the least-squares fits to the spectra and the time dependencies of the optimum parameters for electron solvation in D$_2$O. The remarkable feature of Fig. 7S(b) is that the Gaussian width $W_G$ is constant for $t > 1$ ps: the entire spectral evolution can be explained in terms of a relatively slow increase in $E_{max}$ with a time constant of 1 ps (Fig. 7S(a)) and a concomitant decrease in $v$ (from 2.5 at 1 ps to the equilibrium value of 1.95, with a time constant of 0.6 ps (Fig. 7S(c)). In this model, the spectral evolution during the electron solvation is regarded as the slow spectral shift accompanied by an increase in the curvature of the spectrum at the top. Although the use of eq. (3) with constant $W_L$ and variable $v$ yields as good quality fits as the previously considered model with variable $W_L$ and constant $v$ (and finds equal support in the data on the temperature evolution of $e^-_{aq}$ spectra), entirely different dynamics for $E_{max}$ were obtained from the least-squares fits to the same data set (compare Figs. 6 and 7S). This example illustrates the maxim made in the Introduction: the extraction of $E_{max}$ strongly depends on the spectral template used to fit the data. However, even with this different functional form, the optimum parameters shown in Fig. 7S are inconsistent with the "temperature jump" model: [15] to obtain $v \approx 2.3$ at $t = 1$ ps, a temperature jump of 250 K would be needed (Figs. 1S(a)), whereas the position of $E_{max}$ at this delay time corresponds to a modest temperature jump of 40 K. We conclude that, regardless of how the hydrated electron spectrum is parameterized, the "temperature jump" model is not supported by our data.

**Appendix B. Instruction for the retrieval of kinetic data from the supplied ascii file.**

Included in this Supplement is a 288 Kbyte ASCII file named **H2O_D2O_traces.txt** containing the kinetic traces for light and heavy water. These kinetics were sampled with a time step of 50 fs out to 5-7 ps. We also have a nearly complete set of such kinetics sampled with a step of 500 fs out to 25 ps. The data are formatted as an Igor text wave (suitable for a graphing/analysis program Igor Pro 4.0 or 5.0 from WaveMetrics, Inc.; http://www.wavemetrics.com). It is simple to read these data using other popular graphics programs. Each kinetic trace is tagged by its **<id>** label which consists of the wavelength $\lambda$ in nanometers and identifier "H" (for H$_2$O) or "D" (D$_2$O) at the end (e.g., t1150H is the array of delay times for the electron in light water at 1150 nm). The three columns for each section of the record (that starts with **BEGIN** and ends with **END**) that corresponds to a given wavelength are

t<id>:    delay time $t$ in picoseconds

od<id>:    photoinduced optical density $S(\lambda,t)$ (see section 2).

dod<id>:    ±95% confidence limits for the quantity above.

The format of the data file is

```
IGOR
WAVES/D    t<id1> od<id1>      dod<id1>
BEGIN
<time t, ps>  <normalized S(λ,t)>  <± 95% confidence limits>
....
END
X SetScale/P x 0,1,"", t<id1>; SetScale y 0,0,"", t<id1>
X SetScale/P x 0,1,"", od<id1>; SetScale y 0,0,"", od<id1>
X SetScale/P x 0,1,"", dod<id1>; SetScale y 0,0,"", dod<id1>

WAVES/D    t<id2> od<id2>dod<id2>
BEGIN
...
END
X SetScale/P x 0,1,"", t<id2>; SetScale y 0,0,"", t<id2>
X SetScale/P x 0,1,"", od<id2>; SetScale y 0,0,"", od<id2>
X SetScale/P x 0,1,"", dod<id2>; SetScale y 0,0,"", dod<id2>

.....
etc.
```

For further instruction concerning the data, contact I. A. Shkrob at **shkrob@anl.gov**.

**Table 1S.**

**A comparison between the normalized spectra of hydrated electron in $H_2O$ at 24°C.**

| $\lambda$, μm | this work | ref. 38 [b)] | ref. 26 |
|---|---|---|---|
| 0.5 | 0.290 | 0.32 | 0.293 |
| 0.6 | 0.656 | 0.699 | 0.647 |
| 0.7 [a)] | 1 | 1 | 1 |
| 0.8 | 0.902 | 0.841 | 0.871 |
| 0.9 | 0.572 | 0.502 | 0.57 |
| 1.0 | 0.313 | 0.261 | 0.295 |
| 1.1 | 0.163 | 0.13 | 0.148 |
| 1.2 | 0.085 | 0.066 | 0.082 |
| 1.3 | 0.045 | 0.034 | 0.042 |
| 1.35 | 0.033 | 0.025 | 0.033 |
| 1.4 | 0.025 | 0.018 | 0.024 |

a) the spectra were normalized at this wavelength.
b) calculated using eqs. (2), (3) and (A1) to (A3).

**Figure captions.**

**Fig. 1S**

Temperature dependencies of (a) the parameter $v$ (eq. (A1)) and (b) the energy of the band maximum $E_{max}$ *(to the left;* eqs. (A2) and (A4)) and Gaussian half-width $W_G/\sqrt{\ln 2}$ *(to the right;* eqs. (A3) and (A5)) for hydrated electron in liquid $H_2O$ *(solid lines)* and $D_2O$ *(dashed lines)*. See Appendix A for more detail.

**Fig. 2S**

*Vide infra:* Normalized TA signal from the hydrated electron generated by 248 nm laser photolysis of ferrocyanide *(filled circles)* and sulfite *(open squares)* in the room-temperature liquid $H_2O$. See section A.2 in Appendix A for more detail. Purple circles indicate that data obtained using a Ge photodiode detector, red circles are for the data obtained using a Si photodiode. The red line is the least-squares fit to eqs. (2) and (3) with $v = 2$ (see section A.2 for the optimum parameters). The green line gives the simulated spectrum obtained using the parameters given by eqs. (A1) to (A5). *Vide supra:* the residuals *(same color coding)*.

**Fig. 3S**

Same as Fig. 2S, for hydrated electron in $D_2O$ (ferrocyanide data only). The open squares and open circles stand for the data obtained using Ge and Si detectors, respectively. The solid line is the fit obtained as explained in section A.2.

**Fig. 4S**

The effect of addition of 1 M perchloric acid to water on the decay kinetics of hydrated electron in liquid room-temperature $H_2O$. The TA signal obtained for the probe wavelength of 1.3 μm is shown in the acidic solution *(open squares)* and neat water *(open circles)*. The initial "spike" from pre-solvated electron is not shown; the kinetic traces were normalized at $t=5$ ps. The vertical bars are 95% confidence limits for each data point. The Gaussian radii of the 200 nm pump and 1300 nm probe beams were 37 μm and 14 μm, respectively. The kinetic traces indicated by the symbols were obtained using 2.3 μJ excitation pulse, the traces indicated by slim vertical bars with caps were obtained at lower power, 1.3 μJ. The two series of data match well, suggesting inefficient cross recombination. The solid line is the exponential function that corresponds to a time constant of 43 ps which is close to the life time of hydrated electron that undergoes reaction with $H_3O^+$ in the bulk ($k = 2.3 \times 10^{10}$ M$^{-1}$ s$^{-1}$, see Table 1S). These kinetic data suggest that at least 90-95% of the $\Delta OD_{1300}$ signal at $t>5$ ps is from the hydrated electron.

**Fig. 5S**

The spectral data of (a) Fig. 4(a) and (b) Fig. 5(a) replotted as a function of the photon energy $E$ of the probe light.

**Fig. 6S**

TA spectra for the electron in $D_2O$ analyzed using eqs. (2) and (3) with the variable parameter $v$ and fixed parameter $W_L$ (i.e., in the same way as the high-temperature spectra analyzed by Bartels et al. [37,38] and discussed in section A.1, eqs. (A1) to (A5)). See section A.3 for more detail. The time windows are given in the color scale to the right of the plot. The vertical bars are 95% confidence limits. The solid lines are least-square fits; the (two lower) dashed lines are guides to the eye. The bold solid (yellow) line is the spectrum $S(E)$ of hydrated electron in $D_2O$. The time dependence of the optimum parameters is given in Fig. 7S.

**Fig. 7S**

Time profiles of the optimum parameters (a) $E_{max}$, (b) $W_G$, and (c) $v$ for the electron in $D_2O$. See section 4.2 and Fig. 6S for more detail.

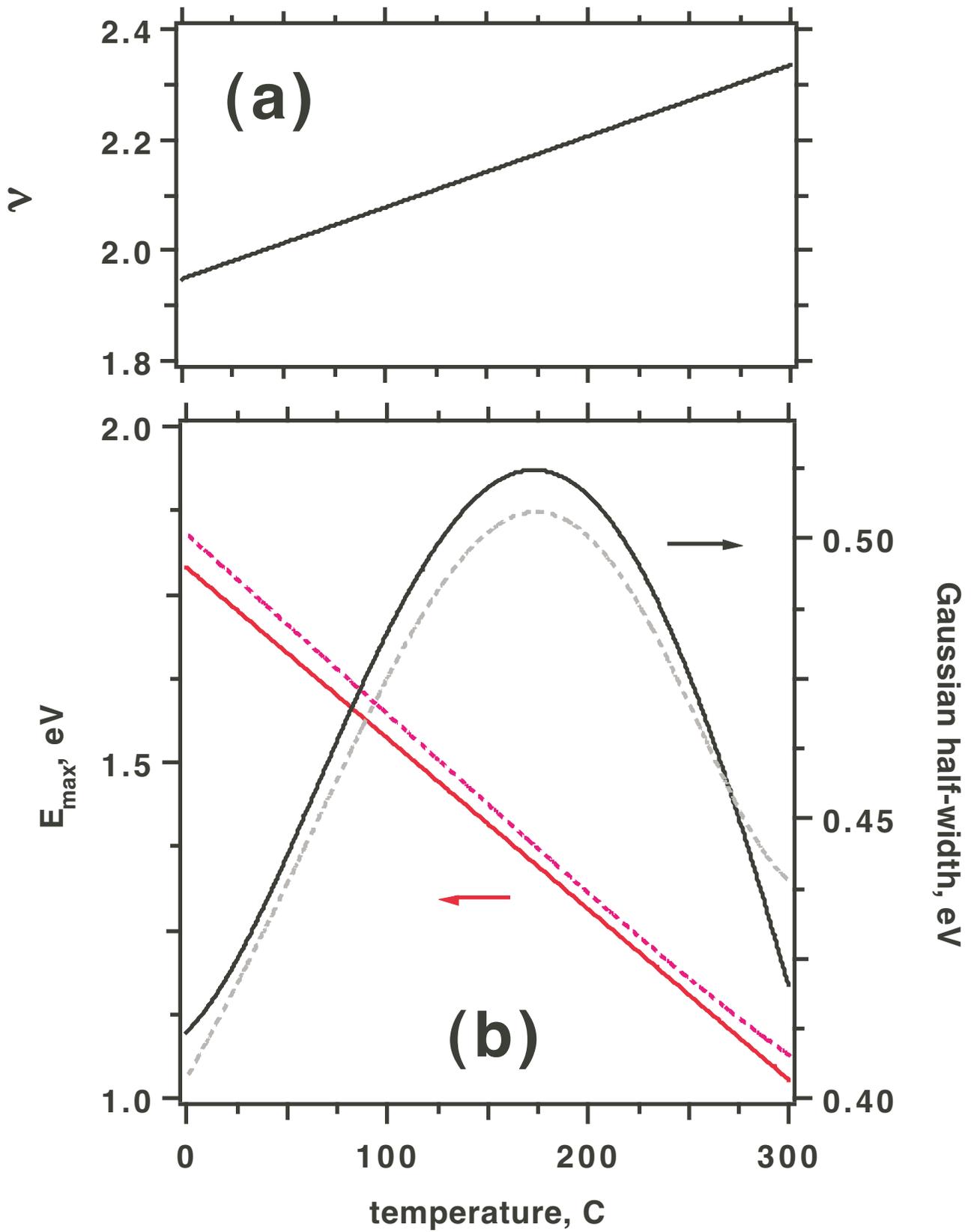

Lian et al., Figure 1S

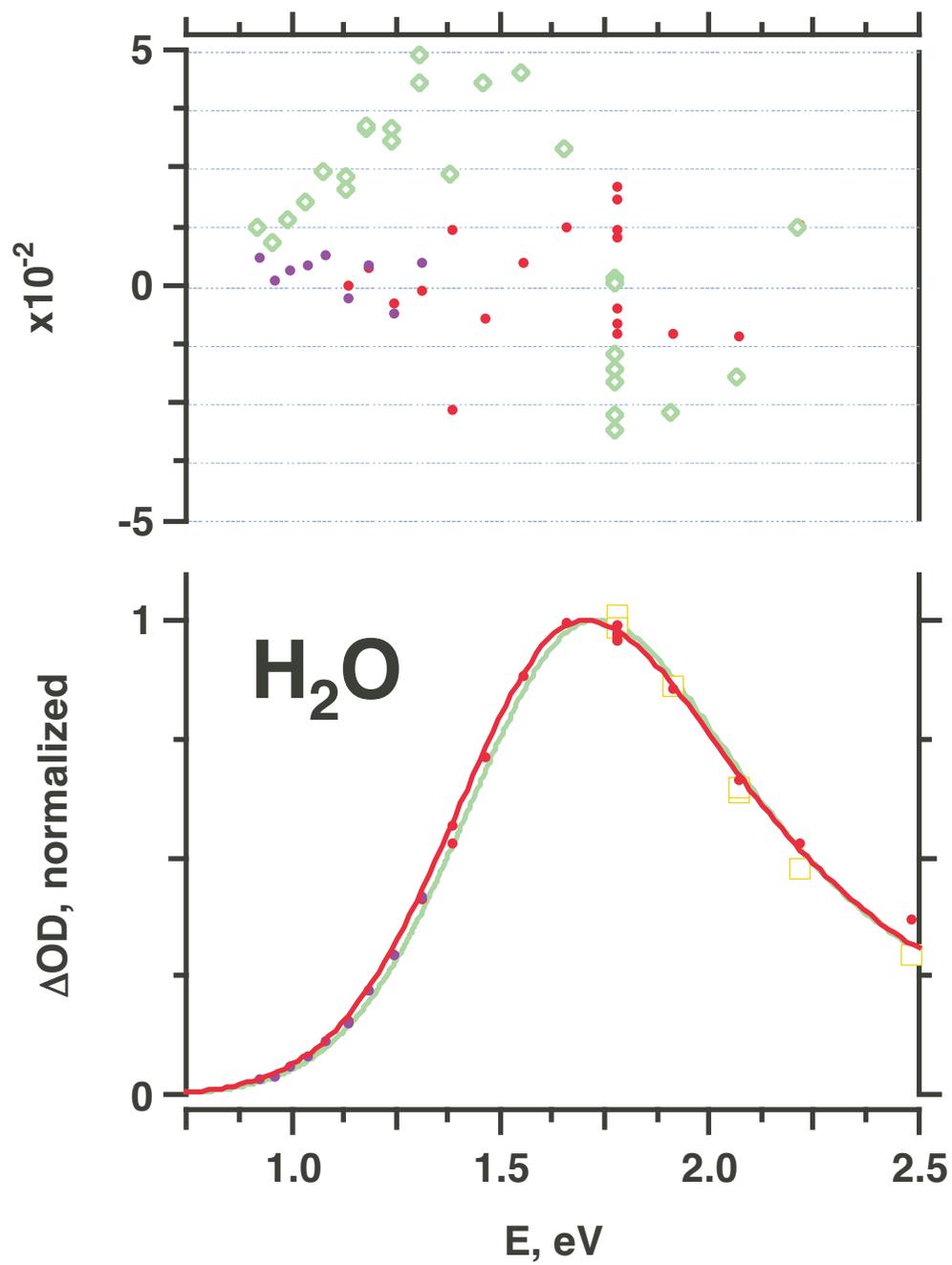

Lian et al., Figure 2S

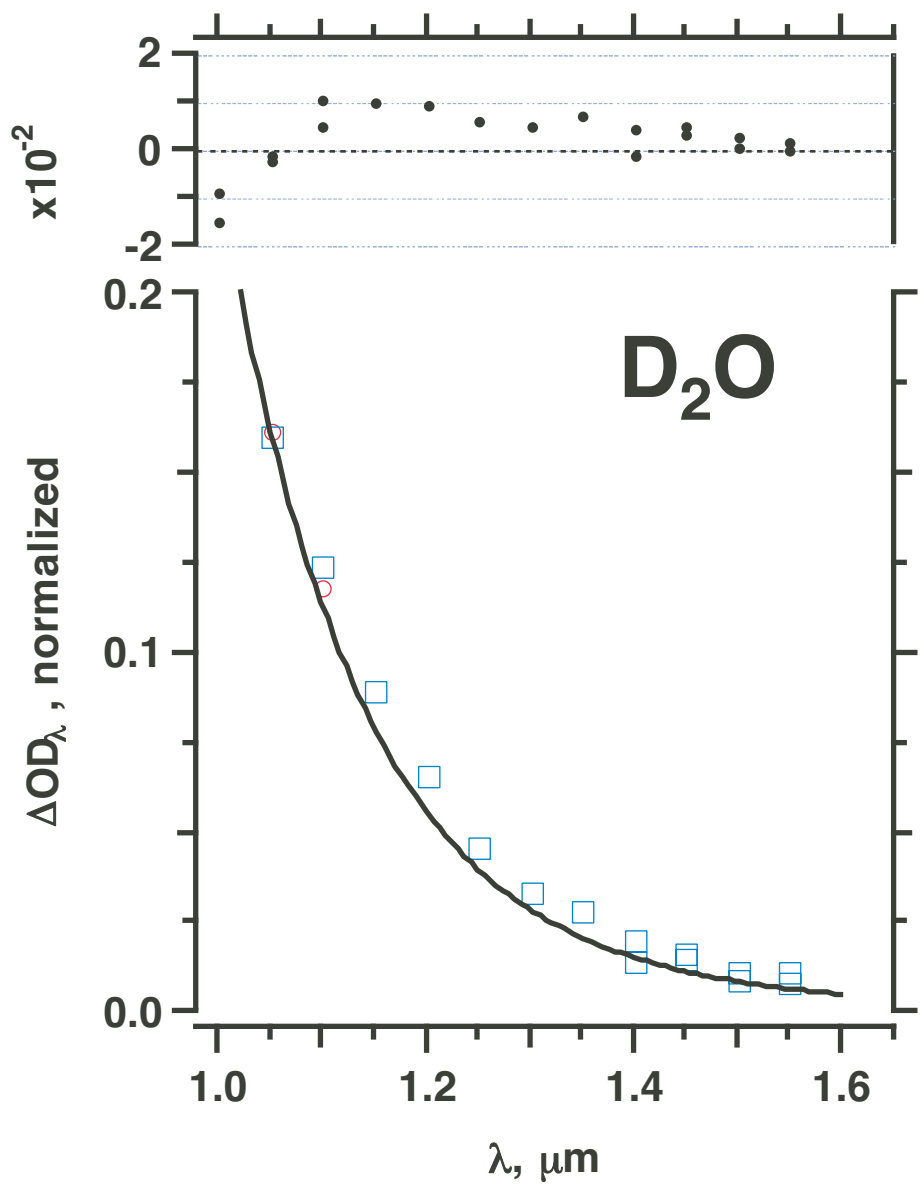

Lian et al., Figure 3S

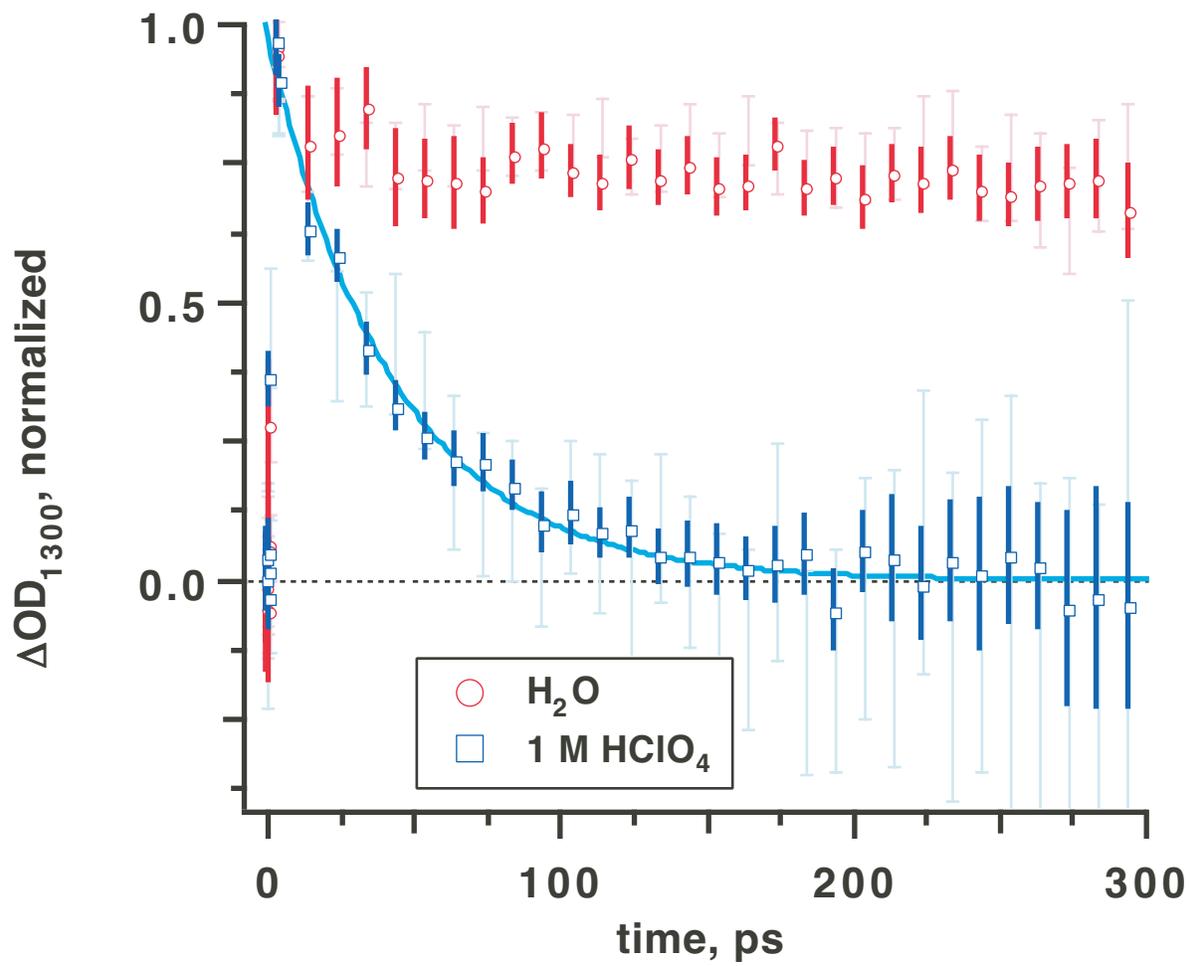

Lian et al., Figure 4S

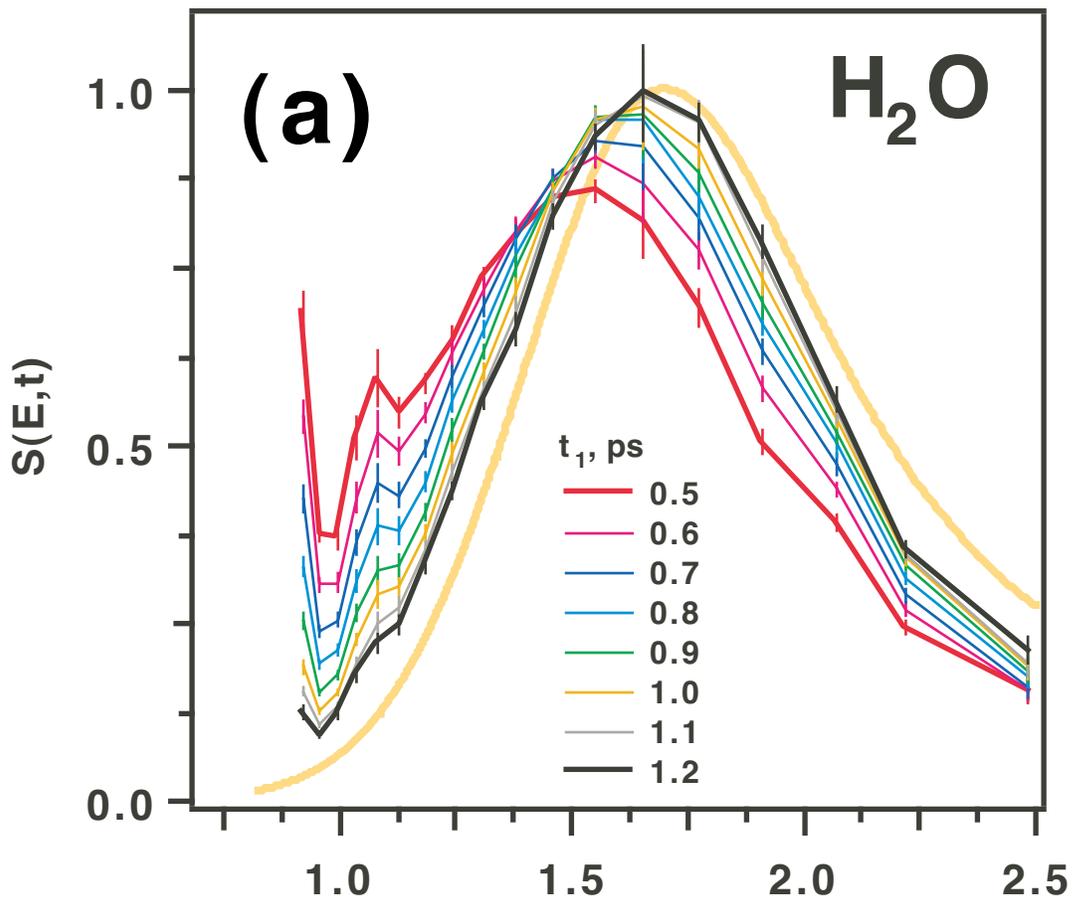
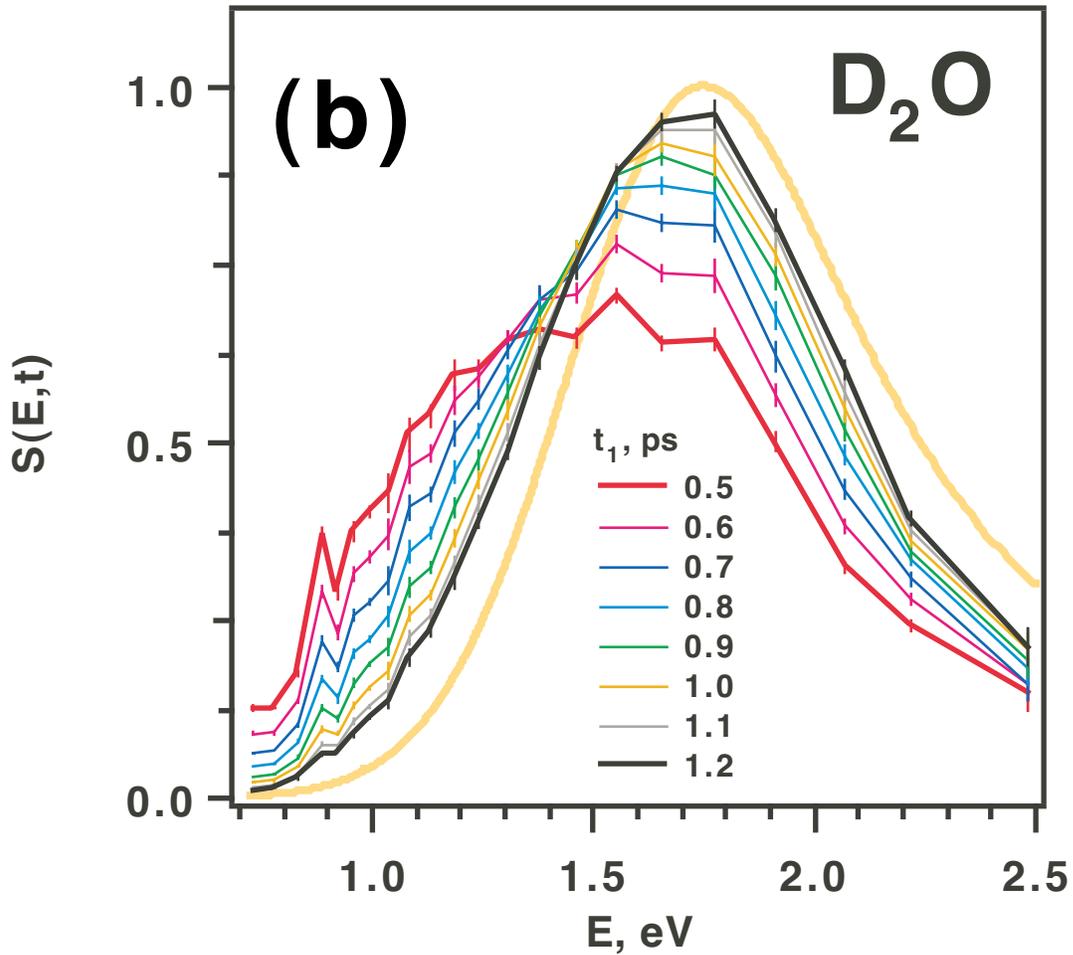

Lian et al., Figure 5S

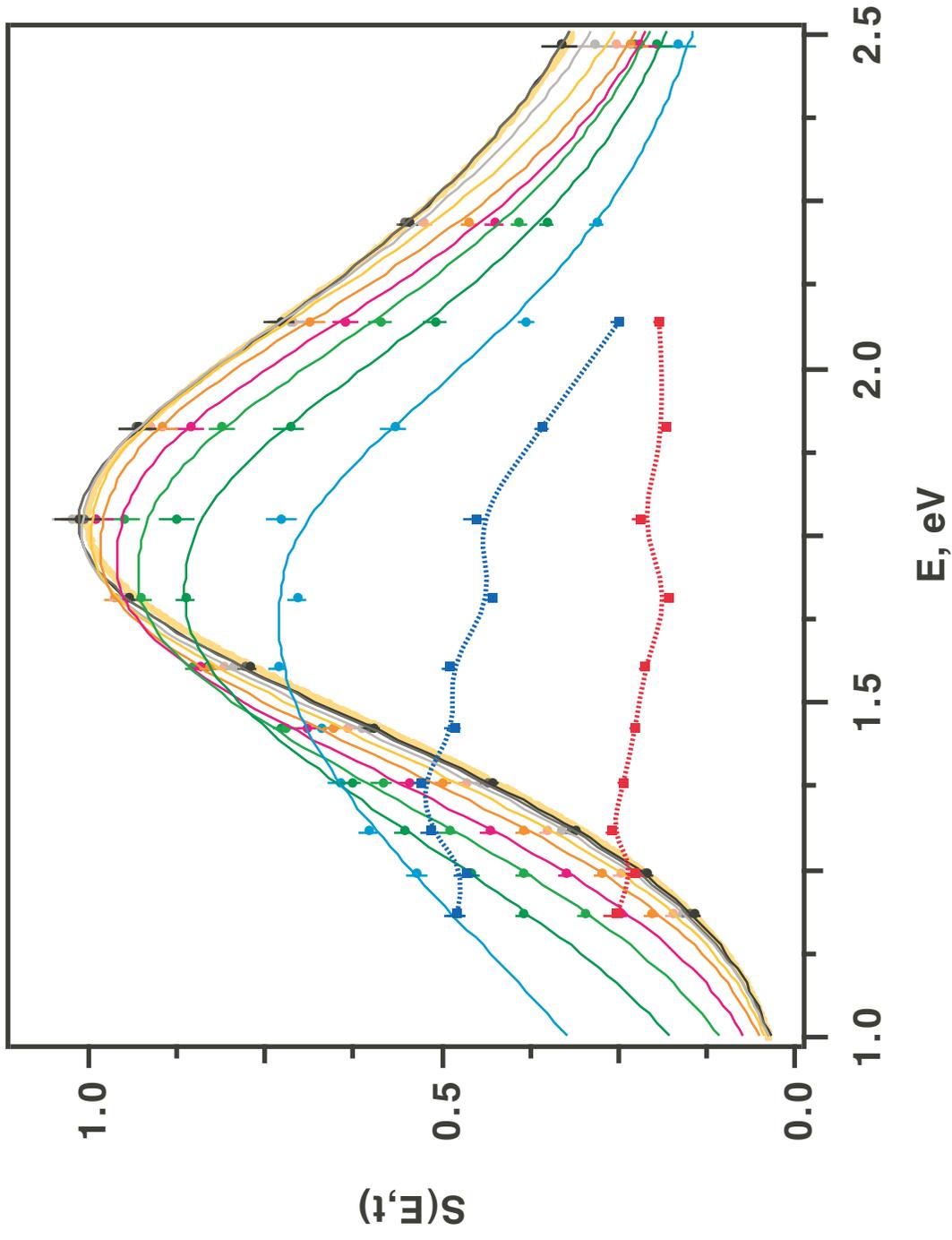

Lian et al., Figure 6S

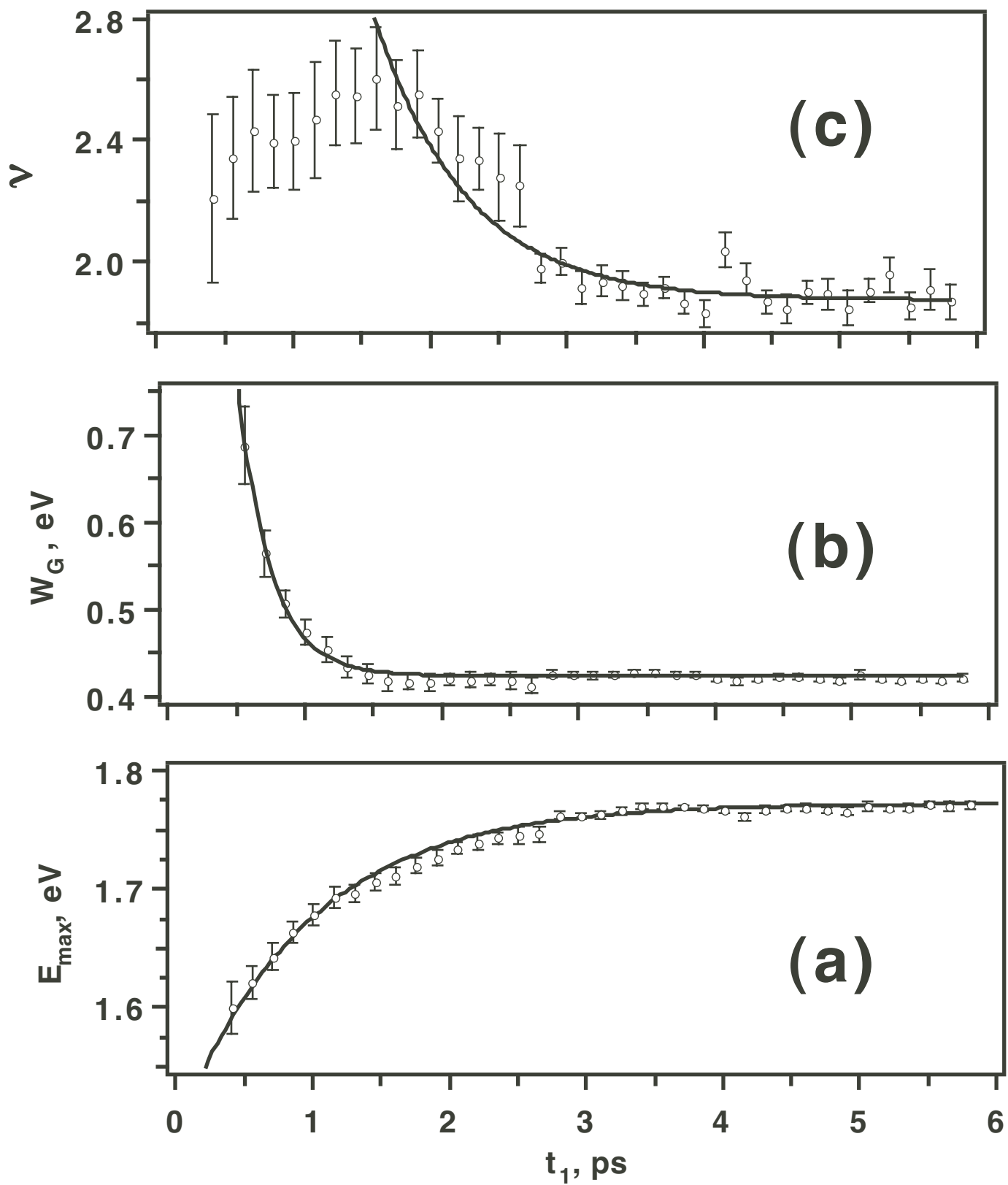

Lian et al., Figure 7S